\documentclass[8.5pt,twoside,twocolumn]{article}
\usepackage[super,sort&compress,comma]{natbib}
\usepackage{mhchem}
\usepackage{times,mathptmx}
\usepackage{sectsty}
\usepackage{balance}

\usepackage{graphicx} 
\usepackage{lastpage}
\usepackage[format=plain,justification=raggedright,singlelinecheck=false,font=small,labelfont=bf,labelsep=space]{caption}
\usepackage{fancyhdr}

\usepackage{amssymb,amsmath,amsfonts}
\graphicspath{{FIGs/}}
\usepackage[mathcal]{eucal}

\begin{document}

\thispagestyle{plain}
\fancypagestyle{plain}{
\renewcommand{\headrulewidth}{1pt}}
\renewcommand{\thefootnote}{\fnsymbol{footnote}}
\renewcommand\footnoterule{\vspace*{1pt}%
\hrule width 3.4in height 0.4pt \vspace*{5pt}}
\setcounter{secnumdepth}{5}

\makeatletter
\def\subsubsection{\@startsection{subsubsection}{3}{10pt}{-1.25ex plus -1ex minus -.1ex}{0ex plus 0ex}{\normalsize\bf}}
\def\paragraph{\@startsection{paragraph}{4}{10pt}{-1.25ex plus -1ex minus -.1ex}{0ex plus 0ex}{\normalsize\textit}}
\renewcommand\@biblabel[1]{#1}
\renewcommand\@makefntext[1]%
{\noindent\makebox[0pt][r]{\@thefnmark\,}#1}
\makeatother
\renewcommand{\figurename}{\small{Fig.}~}
\sectionfont{\large}
\subsectionfont{\normalsize}

\fancyfoot{}
\fancyfoot[RO]{\footnotesize{\sffamily{1--\pageref{LastPage} ~\textbar  \hspace{2pt}\thepage}}}
\fancyfoot[LE]{\footnotesize{\sffamily{\thepage~\textbar\hspace{3.45cm} 1--\pageref{LastPage}}}}
\fancyhead{}
\renewcommand{\headrulewidth}{1pt}
\renewcommand{\footrulewidth}{1pt}
\setlength{\arrayrulewidth}{1pt}
\setlength{\columnsep}{6.5mm}
\setlength\bibsep{1pt}

\twocolumn[
  \begin{@twocolumnfalse}
\noindent\LARGE{\textbf{Interplay between dipole and quadrupole modes of field influence in liquid-crystalline suspensions of ferromagnetic particles
}}

\vspace{0.6cm}

\noindent\large{\textbf{D. V. Makarov,$^{\ast}$ A. N. Zakhlevnykh$^{\ast}$}}\vspace{0.5cm}

\noindent\textit{\small{\textbf{Received Xth XXXXXXXXXX 20XX, Accepted Xth XXXXXXXXX 20XX\newline
First published on the web Xth XXXXXXXXXX 200X}}}

\noindent \textbf{\small{DOI: 10.1039/b000000x}}
\vspace{0.6cm}

\noindent \normalsize{
In the framework of continuum theory we study orientational transitions induced by electric and magnetic fields in ferronematics, \textit{i.e.}, in liquid-crystalline suspensions of ferromagnetic particles. We have shown that in a certain electric field range the magnetic field can induce a sequence of re-entrant orientational transitions in ferronematic layer: nonuniform phase --- uniform phase --- nonuniform phase. This phenomenon is caused by the interplay between the dipole (ferromagnetic) and quadrupole (dielectric and diamagnetic) mechanisms of the field influence on a ferronematic structure. We have found that these re-entrant Freedericksz transitions exhibit tricritical behavior, \textit{i.e.}, they can be of the first or the second order. The character of the transitions depends on a degree of redistribution of magnetic admixture in the sample exposed to uniform magnetic field (magnetic segregation). We demonstrate how electric and magnetic fields can change the order of orientational transitions in ferronematics. We show that electric Freedericksz transitions in ferronematics subjected to magnetic field have no re-entrant nature. Tricritical segregation parameters for the transitions induced by electric or magnetic fields are obtained analytically. We demonstrate the re-entrant behavior of ferronematic  by numerical simulations of the magnetization and optical phase lag.
}
\vspace{0.5cm}
 \end{@twocolumnfalse}
  ]

\section{Introduction}
\footnotetext{\textit{Physics of Phase Transitions Department, Perm State University, Bukirev St. 15, 614990 Perm, Russia.  E-mail: dmakarov@psu.ru; anz@psu.ru}}

It is known, that liquid ferromagnetics do not occur in nature since melting temperature of all known substances is always higher than the so-called Curie temperature at which the long-range order in orientation of magnetic moments of atoms is destroyed and the phase transition to paramagnetic state takes place. For this reason liquid ferromagnetics can only be the artificially fabricated media. First were synthesized the so-called magnetic fluids \cite{Shliomis_1974_UFN} which are colloidal suspensions of magnetic particles in isotropic liquids. These suspensions do not possess ferromagnetic properties since after the external magnetic field is off they do not retain the residual magnetization. Magnetic fluids are inherently paramagnetic and because of their high magnetic susceptibility are often referred to as ``superparamagnetics''.

However, it is possible to synthesize a suspension with ferromagnetic properties on the basis of an anisotropic liquid.  One of such materials is ferronematic \cite{deGennesBook_en, BrochGennes_1970_JPhys} (FN), \textit{i.e.}, low concentrated suspension of anisometric magnetic particles in nematic liquid crystal (NLC).  In order to create a noncompensated ferronematic possessing spontaneous magnetization, we need to add anisometric magnetic particles into the liquid crystal heated above the clearing point and then cool it to liquid crystalline state in an external magnetic field. As a result, the ferronematic will possess magnetization even in the absence of a magnetic field. Due to the liquid crystal matrix FN has good fluidity and anisotropy of physical properties, and embedded in NLC-matrix ferroparticles stipulate a strong magnetic response of the suspension.

In addition to fluidity and high sensitivity to the external fields, ferronematics demonstrate the phenomenon of magnetic admixture redistribution in a uniform magnetic field. Since the magnetic particles are not fixed to the NLC-matrix, they have the opportunity of spatial movement migrating to areas where the sum of their magnetic and orientation energies in liquid crystal matrix is minimal. The phenomenon of magnetic particles accumulation in ``favorable'' parts of the sample under a uniform magnetic field is called the segregation effect. \cite{BrochGennes_1970_JPhys} This effect leads to tricritical behavior both ferrocholesteric--ferronematic transition \cite{ShavkunovZakh_2001_MCLC} and magnetic Freedericksz transition in ferronematics. \cite{MakarovZakh_2010_PRE, ZakhSem_2011_MCLC}

Due to coupling of magnetic particles with liquid crystal molecules there exist two mechanisms of magnetic field effect on the orientational structure of the suspension: the magnetic quadrupole (influence on the liquid-crystal matrix) and magnetic dipole (influence on the magnetic particles). The competition between these mechanisms leads to the threshold Freedericksz-like transition in unbounded ferronematics \cite{Zakhlevn_2004_JMMM} and re-entrant orientational transitions in ferrocholesterics. \cite{ZakhShavkunov_1999_MCLC} An electric field directly acting only on the LC-matrix gives one more mechanism of influence --- the electric quadrupole. The competition between these three mechanisms additionally leads to re-entrant orientational transitions in ferronematics. \cite{MakarovZakh_2012_MCLC} As is shown in Ref. \cite{ZakhSem_2012_ZhTF_EN} the re-entrant transitions are possible in ferronematics subjected to the magnetic field without electric field, but under special bistable coupling conditions on the bounding plates.

In recent years there have been many experimental \cite{Kopcansky_2008_PRE, Kopcansky_2010_JPhysConfSer, Arantes_2010_PhysicsProcedia, Kopcansky_2010_PhysicsProcedia, Ouskova_2010_MCLC, Kopcansky_2010a_JMMM} and theoretical \cite{Zadorozh_2007_MCLC, Zadorozhnii_2010_MCLC, Podoliak_2011_SoftMatter} works on the physical properties of magnetic suspensions based on liquid crystals and the peculiarities of their behavior in external fields. A detailed review of the history of research, classification, the most interesting experimental and theoretical results, as well as numerous applications of liquid-crystalline nanocolloids can be found in Ref. \cite{Garbovskiy_2010_SSP}.

In this paper we study magnetic- and electric-field-induced re-entrant orientational transitions in ferronematics. Taking into account segregation effects leading to the tricritical behavior, \cite{MakarovZakh_2010_PRE} we emphasize on interplay between dipolar (ferromagnetic) and quadrupole (dielectric and diamagnetic) modes of field influence in ferronematics. Our aim is to study the role of these interactions as well as segregation effects in tricritical behavior of re-entrant orientational transitions in ferronematics. We ask ourselves: is it possible to change the order of the Freedericksz transition in ferronematics by external fields?

The rest of the paper is organized as follows. Section~\ref{sec_Equilibrium_equations_Reentrant_phases} introduces a set of equations  describing the orientational structure of ferronematic in electric and magnetic fields taking into account the segregation effect. In Sec.~\ref{sec_Tricritical behavior_Reentrant_phases} we analyze the re-entrant magnetic Freedericksz transitions  in electric field and electric transitions in magnetic field using Landau expansion of ferronematic free energy. Section~\ref{sec_Numerical_results_Reentrant_phases} is devoted to orientational, magnetic and optical properties of ferronematics. Concluding remarks are presented in Sec.~\ref{sec_Conclusion}.

\section{\label{sec_Equilibrium_equations_Reentrant_phases} Orientational equilibrium equations}

\begin{figure}[!htb]
\begin{center}
\includegraphics[width=0.65\linewidth]{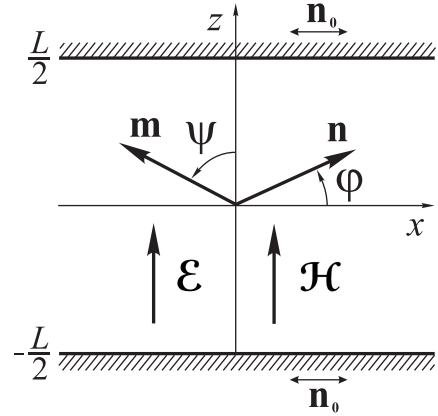}
\caption{Orientation of ferronematic layer in electric $\boldsymbol{\mathcal{E}} $  and magnetic $\boldsymbol{\mathcal{H}} $ fields}
\label{fig1}
\end{center}
\end{figure}

We consider a layer of ferronematic liquid crystal having thickness of $L$, placed between two parallel plates (see Fig.~\ref{fig1}).
Let us introduce a rectangular coordinate system, axis $x$ is directed along the plates, axis $z$  --- perpendicular to the plates; we choose the origin in the middle of the layer. Coupling of the director $\mathbf{n}$ to the layer limiting plates we assume to be rigid (\textit{i.e.}, orientation of the director at the boundary does not change under the influence of the external field) and planar (axis of easy orientation on the layer surface $\mathbf{n}_0$ is parallel to the axis $x$). We apply electric $\boldsymbol{\mathcal{E}}= \textbf{(} 0, \, 0, \, \mathcal{E} \textbf{)}$ and magnetic $\boldsymbol{\mathcal{H}} = \textbf{(} 0, \, 0, \, \mathcal{H} \textbf{)}$ fields orthogonally to the ferronematic layer.

Equilibrium state corresponds to the minimum of free energy functional $\mathcal{F}= \int F_V \, dV$. The bulk free-energy density $F_V$ of ferronematic in electric and magnetic fields under soft surface coupling of magnetic particles to NLC-matrix can be written as follows \cite{BrochGennes_1970_JPhys, BurylRaikh_1995_MCLC2}
\begin{eqnarray} \label{F_FN_layer_Reentrant_phases}
   F_V &=& F_1+F_2+F_3+F_4+F_5+F_6, \\
   F_1 &=& \frac{1}{2} \left[  K_1 (\boldsymbol{\nabla} \cdot
    \mathbf{n})^2 +
  K_2 ( \mathbf{n} \cdot \boldsymbol{\nabla} \times \mathbf{n})^2
   + K_3 ( \mathbf{n} \times \boldsymbol{\nabla}
 \times \mathbf{n})^2 \right], \nonumber \\
   F_2 &=& - \frac{1}{2} \mu_0 \chi_a ( \mathbf{n} \cdot \boldsymbol{\mathcal{H}} )^2,
   \qquad  F_3 = - \mu_0 M_s f \, \mathbf{m} \cdot \boldsymbol{\mathcal{H}} , \nonumber \\
   F_4 &=&  \frac{k_B T}{\nu} \, f  \ln f, \hspace{4.5em} F_5 =
    \frac{w}{d} \,  f ( \mathbf{n} \cdot \mathbf{m})^2, \nonumber \\
   F_6 &=& - \frac{1}{2} \varepsilon_0 \varepsilon_a ( \mathbf{n} \cdot \boldsymbol{\mathcal{E}})^2. \nonumber
\end{eqnarray}
Here $K_1$, $K_2$ and $K_3$ are elastic modules of NLC (Frank constants),   $\chi_a$ is the anisotropy of diamagnetic susceptibility of NLC, $\mu_0$ is the permeability of vacuum, $M_s$ is the saturation magnetization of magnetic particles material, $\varepsilon_0$ is the permittivity of vacuum, $\varepsilon_a$ is the anisotropy of LC permittivity,  $f(\mathbf{r})$  is the local volume fraction of magnetic particles in the suspension, $\mathbf{m}$  is the unit vector of suspension magnetization,  $\nu$  is the volume of a ferroparticle, $d$  is the diameter of a ferroparticle, $k_B$  is the Boltzmann constant, $T$  is the temperature, $w$  is the surface density of coupling energy between nematic LC molecules and the surface of magnetic particles.

The term $F_1$  in expression \eqref{F_FN_layer_Reentrant_phases} represents bulk free-energy density of orientational elastic deformations of the director field (Oseen-Frank potential), $F_2$ is the bulk free-energy density of the magnetic field interaction with nematic matrix (it describes the quadrupole mechanism of the magnetic field influence on FN), $F_3$ is the bulk free-energy density of the magnetic field interaction with magnetic moments of ferroparticles (dipole mechanism of the magnetic field effect on FN), $F_4$ is the contribution of the mixing entropy of the ideal solution of magnetic particles to the bulk free-energy density, $F_5$ is the bulk density of the surface coupling energy of magnetic particles with the director, $F_6$ is the bulk free-energy density of the interaction of electric field with the nematic.

We assume a surface density of coupling energy $w > 0$, so that in the absence of external fields the minimum of free energy corresponds to mutual orthogonal orientation of the director and magnetization $(\mathbf{n} \bot \mathbf{m})$. Let us consider ferronematic with positive anisotropies of diamagnetic susceptibility $\chi_a$  and dielectric permittivity $\varepsilon_a$. Because of initial orthogonal orientation of the director and magnetization  there appears a competition between dipole magnetic  [$\sim \mu_0 M_s f \, \mathbf{m} \cdot \boldsymbol{\mathcal{H}}$] and quadrupole magnetic [$\sim \mu_0 \chi_a (\mathbf{n} \cdot \boldsymbol{\mathcal{H}})^2$] as well as electric [$\sim  \varepsilon_0 \varepsilon_a ( \mathbf{n} \cdot \boldsymbol{\mathcal{E}})^2$] mechanisms of influence in the presence of electric and magnetic fields. Assuming a small volume fraction of ferroparticles in ferronematic ($f \sim 10^{-7}$), but still sufficient for collective behavior of the suspension \cite{Neto_1986_PRA, Matuo_1993_JMMM},  magnetic dipole-dipole interactions can be neglected.

In the geometry under consideration (Fig.~\ref{fig1}) the director $\mathbf{n}$ and unit vector of magnetization $\mathbf{m}$ can be written as
\begin{align} \label{n_m_FN_layer_Reentrant_phases}
 \mathbf{n} &= \textbf{[} \cos\varphi(z), \, 0, \, \sin \varphi(z)\textbf{]}, \nonumber \\
 \mathbf{m} &= \textbf{[} - \sin\psi(z), \, 0, \, \cos \psi(z)\textbf{]}.
\end{align}
To analyze the problem it is convenient to work with dimensionless variables. We choose layer thickness $L$ as a unit of length and introduce the dimensionless coordinate $\widetilde{z}=z/L$. Further on, for simplicity, we omit the tilde sign over the dimensionless variables in the equations. Let us define the dimensionless fields
\begin{align} \label{Fields_FN_layer_Reentrant_phases}
  H = \mathcal{H} L \sqrt{\frac{\mu_0 \chi_a}{K_1}}, \qquad  E = \mathcal{E} L \sqrt{\frac{\varepsilon_0 \varepsilon_a}{K_1}}
\end{align}
 and dimensionless parameters
\begin{align} \label{Parameters_FN_layer_Reentrant_phases}
    b = M_s \overline{f} L\sqrt{\frac{\mu_0}{\chi_a K_1}}, \qquad k = \frac{K_3}{K_1}, \nonumber \\
    \sigma = \frac{w \overline{f} L^2}{K_1 d}, \qquad  \varkappa = \frac{k_B T \overline{f} L^2}{K_1 \nu}.
\end{align}
Here $\overline{f} =  N \nu / V$ is the average volume fraction of magnetic particles in FN.  As a unit of magnetic field $\mathcal{H}$ in \eqref{Fields_FN_layer_Reentrant_phases}  we choose  value $\mathcal{H}_q = L^{-1} \sqrt{K_1/(\mu_0 \chi_a)}$. In this case  at $\mathcal{H} \approx \mathcal{H}_q$ the energy of elastic distortions $F_1$ (Oseen-Frank potential) and diamagnetic term $F_2$ in the free energy of ferronematic \eqref{F_FN_layer_Reentrant_phases} appear to be of the same order. For $\mathcal{H} \gtrsim  \mathcal{H}_q$ orientational distortions are due to diamagnetic anisotropy of LC-matrix (quadrupole mechanism of magnetic field effect on FN). A similar comparison of the elastic $F_1$ and dipole $F_3$ contributions gives another characteristic value of the field  $\mathcal{H}_d = K_1/(\mu_0 M_s \overline{f} L^2)$. For $\mathcal{H} \gtrsim \mathcal{H}_d$ orientational distortions are due to the influence of the magnetic field on the particles (dipole mechanism). Parameter $b = \mathcal{H}_q / \mathcal{H}_d$ is the ratio of these characteristic fields and characterizes the mode of magnetic field effect on FN. \cite{ZakhSosnin_1995_JMMM} For $b > 1$ $(\mathcal{H}_q > \mathcal{H}_d)$ distortion of FN orientational structure in weak fields is caused by the dipole mechanism, whereas for $b < 1$ $(\mathcal{H}_q < \mathcal{H}_d)$ --- quadrupole mechanism. Change of the mode of influence from dipole to quadrupole (and \textit{vice versa}) takes place in the fields for which the contributions  $F_2$ and $F_3$ into the free energy are of the same order, \textit{i.e.}, at $\mathcal{H} \approx \mathcal{H}_0 = M_s \overline{f} / \chi_a$.

As a unit of  electric field $\mathcal{E}$ in \eqref{Fields_FN_layer_Reentrant_phases} we have  chosen the value $\mathcal{E}_q = L^{-1} \sqrt{K_1/(\varepsilon_0 \varepsilon_a)}$, which gives the characteristic value of Freedericksz transition field in nematic. It is determined from the comparison of elastic  $F_1$  and dielectric $F_6$  contributions into free energy \eqref{F_FN_layer_Reentrant_phases}.

We have defined the so-called segregation parameter \cite{ZakhSosnin_1995_JMMM} $\varkappa = (L/ \lambda)^2$ which is the square of the ratio of two characteristic lengths: layer thickness $L$ and segregation length $\lambda = \left(\nu K_1 / k_B T \overline{f}\right)^{1/2}$. The last one characterizes the magnitude of concentrational stratification area in FN. For $\varkappa \gg 1$ the distribution of magnetic particles in a ferronematic layer is close to uniform because characteristic dimensions of the area where concentrational redistribution takes place becomes small in comparison with the thickness of the layer. For $\varkappa \lesssim 1$ the nonuniformity of magnetic particles distribution in a layer becomes significant. Moreover, we introduce dimensionless coupling energy of magnetic particles with the NLC-matrix $\sigma$ and the anisotropy coefficient of orientational elasticity $k$.

It has been already noted, that FN has two mechanisms of response to the applied magnetic field. The first one ($F_3$), linear to the field $H$, stipulates the orientational behavior of FN in weak magnetic fields. This contribution describes the influence of an external magnetic field directly on magnetic moments of ferroparticles, and due to coupling $F_5$ indirectly on the nematic matrix. The second mechanism ($F_2$), quadratic to the field $H$, corresponds to the influence on a diamagnetic LC-matrix and, via it, in accordance with $F_5$, --- to the magnetic moments of ferroparticles. However, in some cases it is necessary to separate the influence onto magnetic particles and onto the LC-matrix. This can be done by placing FN into the electric field, the influence of which on LC-matrix is described by term $F_6$.

Orientational part of the full free energy of a ferronematic $\mathcal{F}$ \eqref{F_FN_layer_Reentrant_phases} in dimensionless form can be written as
\begin{multline} \label{F_FN_layer_dimensionless_Reentrant_phases}
   \widetilde{F} = \frac{L}{K_1 S} \mathcal{F} = \int_{-1/2}^{1/2}
     \left[\frac{1}{2} \mathcal{K}(\varphi) \left(\frac{d \varphi}{dz}\right)^{2} - b H g \cos \psi -   \right.  \\
    \left. - \frac{1}{2} \left(H^2 + E^2 \right) \sin^{2} \varphi + \varkappa g \ln g +  \sigma g \sin^{2} (\varphi - \psi) \right] dz,
\end{multline}
where $S$ is the surface area of the plates limiting FN layer, also we introduced functions
\begin{align} \label{g_function_FN_layer_Reentrant_phases}
  g(z) &= f(z) / \overline{f}, \\
  \mathcal{K}(\varphi) &= \cos^{2} \varphi+ k \sin^{2} \varphi.
\label{K_function_FN_layer_Reentrant_phases}
\end{align}
Here $g(z)$ is the reduced volume fraction of magnetic particles in FN.

Free energy \eqref{F_FN_layer_dimensionless_Reentrant_phases} is a functional over functions $\varphi(z)$, $\psi(z)$ and $g(z)$. Minimization of the free energy \eqref{F_FN_layer_dimensionless_Reentrant_phases}  with respect to  $\varphi(z)$ and $\psi(z)$ gives equations determining orientation angles of the director and magnetization:
\begin{eqnarray}
    \mathcal{K}(\varphi) \, \varphi'' + \displaystyle \frac{1}{2}\frac{d
    \mathcal{K}(\varphi)}{d\varphi} \, {(\varphi')}^2  + \frac{1}{2} \, \left(H^2 + E^2 \right) \sin 2\varphi  \nonumber \\
    - \sigma g \sin 2(\varphi - \psi)  = 0, \label{Eq_System_A_FN_layer_Reentrant_phases}\\
     b H \sin \psi - \sigma \sin 2(\varphi - \psi) = 0, \label{Eq_System_B_FN_layer_Reentrant_phases}
\end{eqnarray}
with the boundary conditions
\begin{equation}\label{boundary_conditions1_FN_layer_Reentrant_phases}
\varphi\left(-1/2\right) = \varphi\left(1/2\right) = 0,
\end{equation}
corresponding to the rigid planar coupling of the director on the surfaces $\mathbf{n}|_{z=\pm1/2} = (1, 0, 0)$. Hereafter the prime denotes the derivative with respect to the dimensionless coordinate $z$.

Equilibrium distribution of magnetic particles over the FN layer is described by minimising the free energy functional \eqref{F_FN_layer_dimensionless_Reentrant_phases} with respect to $g(z)$ under  the condition of constant number of particles in a suspension $\int f dV = N\nu$:
\begin{equation}
     g =  Q \, \exp{\left\{\displaystyle \frac{b H}{\varkappa} \cos \psi  - \frac{\sigma}{\varkappa} \sin^2 (\varphi - \psi)\right\}},
      \label{Eq_System_C_FN_layer_Reentrant_phases}
\end{equation}
where
\begin{equation} \label{Q_function_FN_layer_Reentrant_phases}
Q^{-1} =  \int_{-1/2}^{1/2} \exp{\left\{\displaystyle \frac{b H}{\varkappa} \cos \psi  - \frac{\sigma}{\varkappa} \sin^2 (\varphi - \psi)\right\}}\, d z .
\end{equation}

Note that the so-called coupling equation \eqref{Eq_System_B_FN_layer_Reentrant_phases} determines mutual orientation of the director and magnetization, \cite{BurylRaikh_1995_MCLC2} whereas eqn \eqref{Eq_System_C_FN_layer_Reentrant_phases} describes the segregation effect, \cite{BrochGennes_1970_JPhys} consisting in magnetic particles concentration growth in those parts of the sample where the sum of their magnetic energy in the field $\boldsymbol{\mathcal{H}}$ and orientational energy in LC-matrix is minimal. For $\varkappa \gg 1$ magnetic segregation effects can be neglected. In this case it is obvious from eqns \eqref{Eq_System_C_FN_layer_Reentrant_phases} and \eqref{Q_function_FN_layer_Reentrant_phases} that normalization integral $Q \rightarrow 1$, whereas volume fraction of magnetic particles $g(z) \rightarrow 1$, \textit{i.e.}, $f(z) \rightarrow \overline{f}$.

Let us make estimates of the dimensionless variables \eqref{Parameters_FN_layer_Reentrant_phases} using typical values of material parameters for nematic liquid crystals \cite{deGennesBook_en, BlinovBook_1994_en} and magnetic particles \cite{MakarovZakh_2010_PRE}:  $\chi_a = 2.1 \times 10^{-6}$, $K_1 = 6.4 \times 10^{-12} \, \textrm{N}$, $K_3 = 1.0 \times 10^{-11} \, \textrm{N}$, $T = 298 \, \textrm{K}$, $\bar{f} = 2.0 \times 10^{-7}$, $M_s = 5 \times 10^{5} \, \textrm{A} \textrm{m}^{-1}$, $w = 10^{-6} - 10^{-4} \, \textrm{N} \textrm{m}^{-1}$, $d = 7.5 \times 10^{-8} \, \textrm{m}$, $\nu = 8.8 \times 10^{-22} \, \textrm{m}^{3}$, and taking layer thickness $L = 250 \,  \textrm{$\mu$m}$, we find $k \approx 1$, $\gamma \approx 1$, $\sigma \approx 10^{-2} - 1$, $b \approx 10$ and $\varkappa \approx 10^{-2}$. As can be seen from these estimates, the smallness of $\varkappa$ testifies to the importance of segregation effects in the problem under consideration.

Equations \eqref{Eq_System_A_FN_layer_Reentrant_phases}, \eqref{Eq_System_B_FN_layer_Reentrant_phases},  and \eqref{Eq_System_C_FN_layer_Reentrant_phases} with boundary conditions \eqref{boundary_conditions1_FN_layer_Reentrant_phases} admits a uniform solution $\varphi(z) \equiv  \psi(z) \equiv 0$ corresponding to the planar ($\mathbf{n} = \mathbf{n}_0$) texture of FN with orthogonal ($\mathbf{n} \bot \mathbf{m}$) orientation of the director and magnetization. At the same time they determine also nonuniform orientations of the director and magnetization of FN under mutual influence of electric and magnetic fields. Let us consider nonuniform solutions  for the orientational angles of the director and magnetization. At first we multiply eqn \eqref{Eq_System_A_FN_layer_Reentrant_phases} by $\varphi'$ and eqn \eqref{Eq_System_B_FN_layer_Reentrant_phases} by  $g \psi'$, then we subtract the second equation from the first. As a result we obtain
\begin{equation}\label{Int1_FN_layer_Reentrant_phases}
      \frac{d}{dz} \biggl[ \mathcal{K}(\varphi) \,
       {\left(\varphi'\right) }^2 -  \left(H^2 + E^2 \right) \cos^2 \varphi + 2 \varkappa  g \biggr] = 0.
\end{equation}
Taking into consideration that in the middle of the layer the deviation of the director from the axis of easy orientation is maximal, \textit{i.e.}, $\varphi' = 0$ at $z=0$, the first integral of eqn  \eqref{Int1_FN_layer_Reentrant_phases} takes the form
\begin{equation}\label{Int2_FN_layer_Reentrant_phases}
   \mathcal{K}(\varphi) \, {\left( \varphi' \right) }^2 =
    \left(H^2 + E^2 \right) \left( \cos^2 \varphi - \cos^2 \varphi_0 \right)
      + 2 \varkappa  \left(g_0 - g\right) .
\end{equation}
Here $g_0 \equiv g(\varphi_0, \psi_0)$ is the distribution function of magnetic particles, $\varphi_0 \equiv \varphi(0)$ and $\psi_0 \equiv \psi(0)$ are angles of director and magnetization orientation in the middle of the layer, respectively.

Integration of eqn \eqref{Int2_FN_layer_Reentrant_phases}  for $z>0$  with boundary conditions \eqref{boundary_conditions1_FN_layer_Reentrant_phases} gives
\begin{equation}\label{Int_Eq_z_FN_layer_Reentrant_phases}
  \int_{0}^{\varphi(z)}\sqrt{\mathcal{R}(\varphi, \psi)} \, d\varphi = \frac{1}{2} - z,
\end{equation}
where
\begin{equation} \label{R_function_FN_layer_Reentrant_phases}
\mathcal{R} (\varphi, \psi) =
\frac{\cos^{2} \varphi+ k \sin^{2} \varphi}{\left(H^2 + E^2 \right) \left( \cos^2\varphi  -
\cos^2 \varphi_0 \right) + 2 \varkappa \left(g_0 - g \right)}.
\end{equation}
We choose plus sign in eqn \eqref{Int_Eq_z_FN_layer_Reentrant_phases},  that corresponds to the positive (counterclockwise rotation) values of the angle of the director orientation.

In the middle of the layer ($z=0$) the angle $\varphi$ has maximum value $\varphi_0$, which is determined by the following equation, arising from \eqref{Int_Eq_z_FN_layer_Reentrant_phases}
\begin{equation} \label{Full_Int_Eqa_FN_layer_Reentrant_phases}
\int_{0}^{\varphi_0}\sqrt{\mathcal{R}(\varphi, \psi)} \, d\varphi = \frac{1}{2}.
\end{equation}
Transforming the normalization integral  $Q$ \eqref{Q_function_FN_layer_Reentrant_phases} in eqn \eqref{Eq_System_C_FN_layer_Reentrant_phases} with the help of expression \eqref{Int2_FN_layer_Reentrant_phases},  we get the equation for the distribution function $g(z)$,
\begin{equation} \label{Full_Int_Eqb_FN_layer_Reentrant_phases}
\int_{0}^{\varphi_0}g(\varphi, \psi) \, \sqrt{\mathcal{R}(\varphi, \psi)} \, d\varphi = \frac{1}{2}.
\end{equation}
Thus,  eqns \eqref{Int_Eq_z_FN_layer_Reentrant_phases}--\eqref{Full_Int_Eqb_FN_layer_Reentrant_phases} and coupling equation \eqref{Eq_System_B_FN_layer_Reentrant_phases} determine the angles of orientation of the director $\varphi$ and magnetization $\psi$, and distribution function $g(z)=f(z) / \overline{f}$  of particles as functions of magnetic $H$ and electric $E$ fields and material parameters of ferronematic.

\section{Reentrant Freedericksz transitions in ferronematics} \label{sec_Tricritical behavior_Reentrant_phases}

Equations \eqref{Eq_System_A_FN_layer_Reentrant_phases}, \eqref{Eq_System_B_FN_layer_Reentrant_phases}, and \eqref{Eq_System_C_FN_layer_Reentrant_phases} with boundary conditions \eqref{boundary_conditions1_FN_layer_Reentrant_phases} admit a uniform solution $\varphi(z) = \psi(z) = 0$, corresponding to the planar ($\mathbf{n} = \mathbf{n}_0$) texture of FN with  orthogonal ($\mathbf{n} \bot \mathbf{m}$) orientation of the director and magnetization. However, this solution becomes unstable and is replaced by nonuniform solution when the external field (electric or magnetic) reaches a threshold value. Such transition from uniform to nonuniform state is called the Freedericksz transition \cite{deGennesBook_en}. It is the result of competition of surface forces that orient the liquid crystal on the layer boundary, forces of orientational elasticity and external force fields. In the considered geometry (Fig.~\ref{fig1}) the Freedericksz transition in FN layer can be caused by both magnetic and electric fields. Let us consider first the magnetic-field induced nonuniform orientation of the director  at a given electric field.

\subsection{Magnetic Freedericksz transitions in an electric field}

Close to the Freedericksz transition, angles $\varphi(z)$ and $\psi(z)$ are small, so in the lowest order expansion of eqns  \eqref{Eq_System_A_FN_layer_Reentrant_phases}, \eqref{Eq_System_B_FN_layer_Reentrant_phases}, and \eqref{Eq_System_C_FN_layer_Reentrant_phases} with boundary conditions \eqref{boundary_conditions1_FN_layer_Reentrant_phases} we find
 \begin{equation}\label{PhiPsi_FN_layer_Reentrant_phases}
\varphi(z)=\varphi_0 \cos\pi z, \quad \psi(z) = s_H \varphi(z), \quad s_H = 2 \sigma/(2 \sigma + b H_c),
\end{equation}
where $\varphi_0 \ll 1$, and critical magnetic field $H_c$ is determined by equation \cite{MakarovZakh_2012_MCLC}
 \begin{equation}\label{h_c_FN_layer_Reentrant_phases}
H_{c}^2 = \pi^2 - E^2 +  s_H b H_c.
\end{equation}
At zero magnetic field, $H=0$, it gives the threshold value of the electric field $E_c$ in FN which coincides with the Freedericksz field $E_c^{LC}=\pi$ in pure nematic. \cite{deGennesBook_en} At zero electric field, $E=0$, eqn \eqref{h_c_FN_layer_Reentrant_phases} reduces to the expression obtained in Ref. \cite{MakarovZakh_2010_PRE}.

Free energy \eqref{F_FN_layer_dimensionless_Reentrant_phases}  can be expanded as a power series in the  $\varphi(z)$ and $\psi(z)$ \eqref{PhiPsi_FN_layer_Reentrant_phases}. The fourth-order expansion in the Landau form gives
\begin{equation} \label{F_criticalH_FN_layer_Reentrant_phases}
 \widetilde{F}  = \widetilde{F}_0  + \frac{\alpha_H}{2} \left(H_c - H\right)  \varphi_0^{2} +\frac{\beta_H}{4} \varphi_0^{4} + \dots,
\end{equation}
where
\begin{align*}
\widetilde{F}_0 &= -b H, \quad \alpha_H = \displaystyle \frac{1}{2H_c}\left[H^2_c (2-s_H)+s_H (\pi^2-E^2)\right], \\
 \beta_H &= \frac{1}{16 \varkappa} \left[3b H_c s_H^2 \left(2-s_H \right)^2  +4\pi^2 k\right] \left(\varkappa-\varkappa_H^{*}\right).
\end{align*}
Here we introduce
\begin{equation} \label{kappaH_critical_FN_layer_Reentrant_phases}
 \varkappa_H^{*} = \frac{\left[H^{2}_c + E^2 - \pi^2\right]^2}{3 b H_c s_H^2  \left(2-s_H\right)^2  +4\pi^2 k}.
\end{equation}

The director rotation angle is defined by minimizing of free energy \eqref{F_criticalH_FN_layer_Reentrant_phases} with respect to the $\varphi_0$
\begin{equation} \label{phi0_close_TKT_FN_layer_Reentrant_phases}
 \varphi_0 = \pm \sqrt{\gamma_H \frac{\,H-H_c}{\varkappa-\varkappa_H^{*}}}, \, \gamma_H= \displaystyle \frac{8\varkappa \left[H^{2}_c (2-s_H) + s_H (\pi^2-E^2)\right]}{3b\left[H_c s_H (2-s_H)\right]^2+4\pi^2 k H_c}.
\end{equation}
As seen from formula \eqref{phi0_close_TKT_FN_layer_Reentrant_phases},  segregation parameter $\varkappa_H^{*}$ corresponds to the tricritical point, in which the character of transition between the uniform and nonuniform phases changes from  the second order to the first. For $\varkappa \geq \varkappa_H^{*}$ (weak segregation)  at $H=H_c$ orientational transition is the second order, whereas for  $\varkappa < \varkappa_H^{*}$ (strong segregation)  the transition becomes the first order. In the latter case, ferronematic exhibits orientational bistability. At zero electric field, $E=0$, formulas \eqref{F_criticalH_FN_layer_Reentrant_phases} -- \eqref{phi0_close_TKT_FN_layer_Reentrant_phases} coincide with expressions, obtained in Ref. \cite{MakarovZakh_2010_PRE}.

\begin{figure}[!htb]
\center{\includegraphics[width=0.65\linewidth]{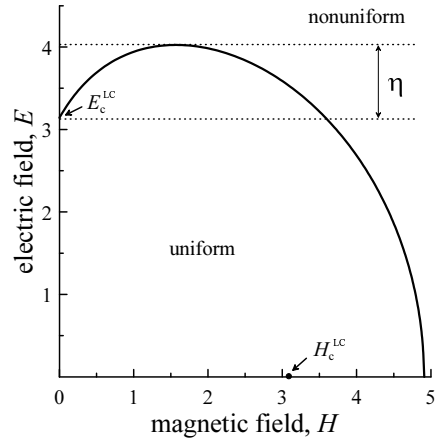}}
\caption{Bifurcation phase diagram of orientational transitions in FN  for  $\sigma=10$ and $b=10$. Here $E^{LC}_c = H^{LC}_c = \pi$ are fields of electric and magnetic Freedericksz transitions in pure nematic, $\eta$ is the width of the region where the re-entrant phases are possible}
\label{fig_Thresholds_EH}
\end{figure}

The bifurcation phase diagram \eqref{h_c_FN_layer_Reentrant_phases} of orientational transitions in FN is plotted  in Fig.~\ref{fig_Thresholds_EH}. In the case of weak magnetic segregation ($\varkappa \geq \varkappa_H^{*}$) the curve $E(H)$  defines the boundary of the second-order Freedericksz transitions in FN induced by the joint action of electric and magnetic fields. The region under the curve $E(H)$ corresponds to the unperturbed state of FN, \textit{i.e.}, uniform planar ($\mathbf{n} = \mathbf{n}_0$)  texture of FN with mutual orthogonal orientation of the director and magnetization ($\mathbf{n} \bot \mathbf{m}$), whereas the region over the curve corresponds to the nonuniform state.

Characteristic feature of the phase diagram (Fig.~\ref{fig_Thresholds_EH}) is the presence of re-entrant phase transitions in FN (nonuniform phase --- uniform phase --- nonuniform phase)  caused by competition between quadrupole electric [$\sim  \varepsilon_0 \varepsilon_a ( \mathbf{n} \cdot \boldsymbol{\mathcal{E}})^2$] and magnetic [$\sim  \mu_0 \chi_a ( \mathbf{n} \cdot \boldsymbol{\mathcal{H}})^2$]  mechanisms of influence on NLC-matrix, and dipole [$ \sim \mu_0  M_s f \, \mathbf{m} \cdot \boldsymbol{\mathcal{H}}$] influence of magnetic field on magnetic moments of ferroparticles. As can be seen from Fig.~\ref{fig_Thresholds_EH}, for  $E_c^{LC} < E \leq E_c^{LC} + \eta$ at $H=0$ ferronematic is in nonuniform phase because $E_c^{LC}$ is a field of electric Freedericksz transition. Application of a magnetic field induces the transition to uniform phase and then with further increase of $H$ uniform phase again is changed to nonuniform. Thus, critical field $H_c$ for a given $E$ has two values: $H_1$ at the transition from nonuniform state to uniform, and $H_2 > H_1$ --- at the re-entrant  transition from uniform state to nonuniform  (Fig.~\ref{fig_Phi0_H}a). In the first case, formula \eqref{phi0_close_TKT_FN_layer_Reentrant_phases} has parameter  $\gamma_H <0$, in the second --- parameter $\gamma_H >0$. For $E > E_c^{LC} + \eta$ Freedericksz transition can not be induced by a magnetic field. The width of the region where the re-entrant phases are possible
 \begin{equation}\label{delta_FN_layer_Reentrant_phases1}
\eta =  \sqrt{\pi^2 - \frac{(q-2\sigma)^4}{9 b^2 q^2} + \frac{2\sigma  (q-2\sigma)^2}{6 q \sigma +  (q - 2\sigma)^2}} - \pi,
\end{equation}
where
\begin{equation*}
q=\sqrt[3]{\sigma^2\left[8\sigma+27 b^2+3 b\sqrt{3(16\sigma + 27 b^2)} \right]},
\end{equation*}
was obtained analytically in Ref. \cite{MakarovZakh_2012_MCLC}. As is seen from \eqref{delta_FN_layer_Reentrant_phases1} the interval of electric field values, admitting re-entrant uniform phase of FN, expands with the increase of coupling energy $\sigma$ and the parameter $b$, and does not depend on segregation degree, which is characterized by the parameter $\varkappa$.

For the magnetic Freedericksz transition in a non-zero electric field, the tricritical value $\varkappa_H^{*}$ of segregation parameter  \eqref{kappaH_critical_FN_layer_Reentrant_phases}  is a function of $E$.
\begin{figure}[!htb]
\begin{minipage}[!htb]{1.0\linewidth}
\center{\includegraphics[width=0.65\linewidth]{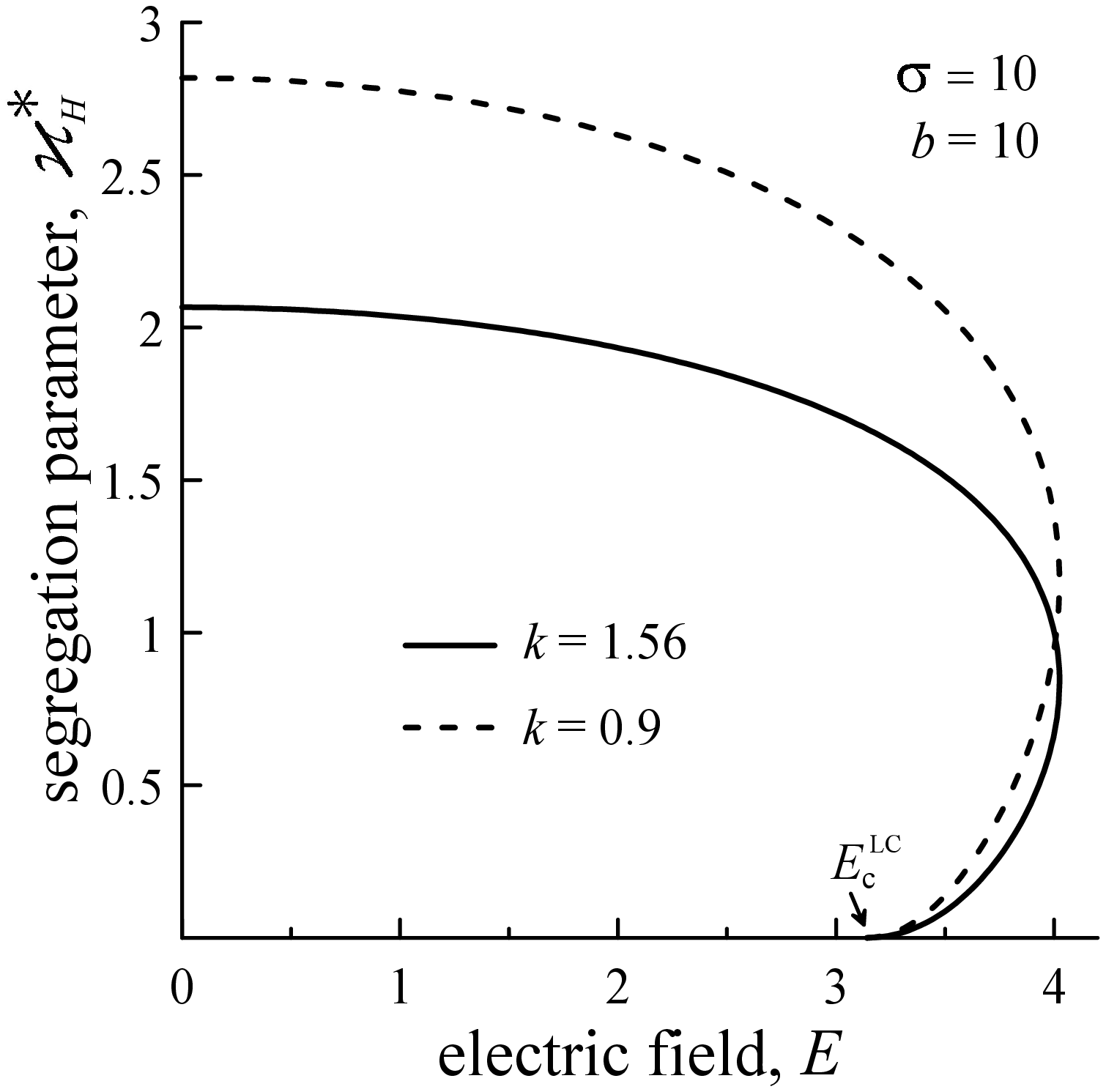} \\ (a)}
\end{minipage}
\vfill
\begin{minipage}[!htb]{1.0\linewidth}
\center{\includegraphics[width=0.65\linewidth]{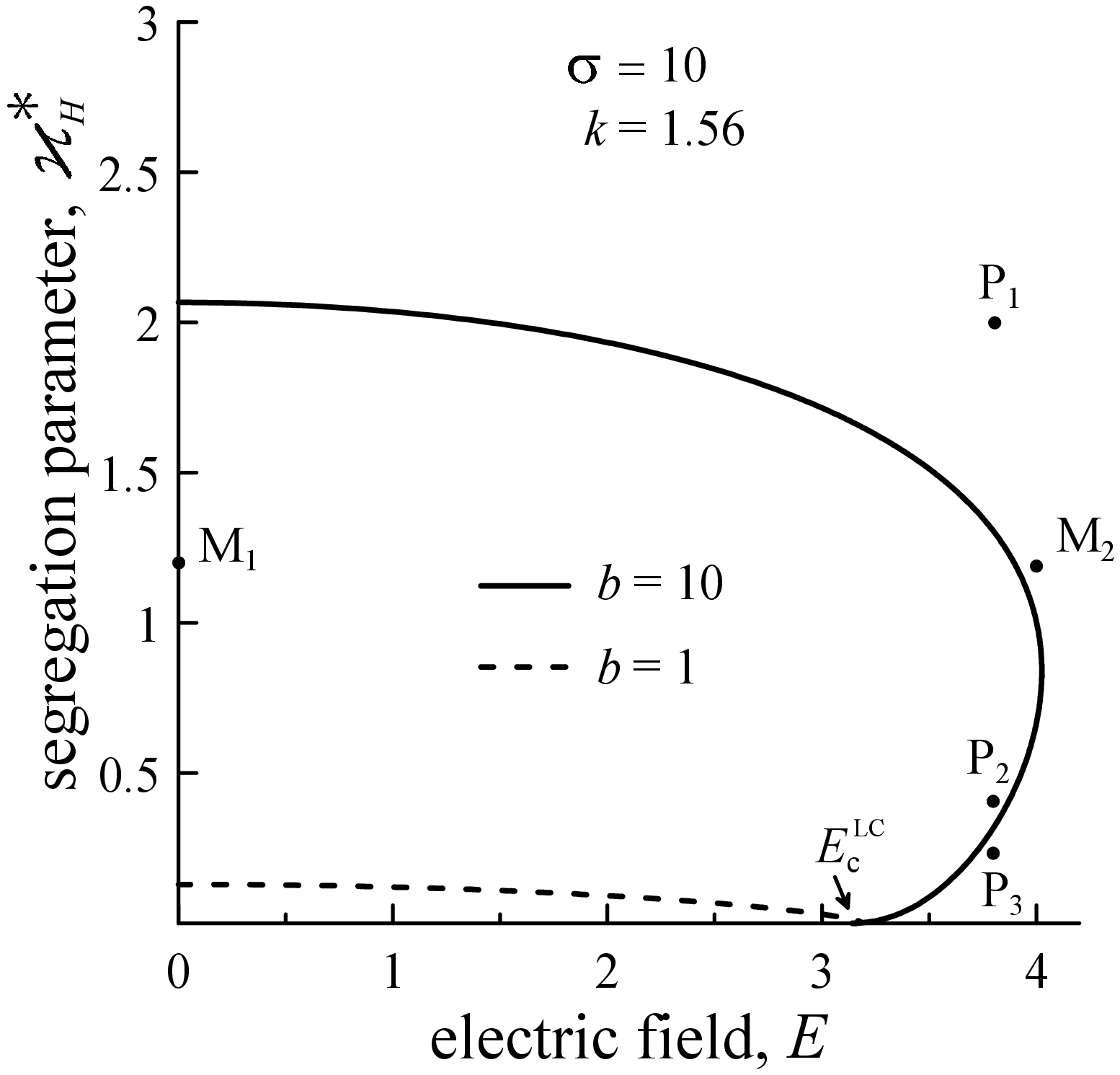} \\ (b)}
\end{minipage}
\caption{Tricritical value of segregation parameter $\varkappa_H^{*}$ as a function of electric field $E$ for different values of (a) orientational elasticity anisotropy constant $k$ and (b) parameter $b$, which characterizes the modes of magnetic field influence on ferronematic;  $M_1=(0, 1.2)$,  $M_2=(4.0, 1.2)$, $P_1=(3.8, 2.0)$, $P_2=(3.8, 0.4)$, $P_3=(3.8, 0.25)$}
\label{fig_kappaH_TKT}
\end{figure}
As the electric field increases the parameter  $\varkappa_H^{*}$ decreases as shown in Fig.~\ref{fig_kappaH_TKT}. Thus at zero electric field the magnetic  Freedericksz  transition in FN is the first-order transition ($\varkappa < \varkappa_H^{*}$). The application of electric field can decrease the value $\varkappa_H^{*}$ so that will be satisfied the inequality $\varkappa > \varkappa_H^{*}$ and magnetic Freedericksz transition becomes of the second order.  Such behavior of the orientational transition is confirmed by numerical simulation of eqns \eqref{Eq_System_B_FN_layer_Reentrant_phases}, \eqref{Int_Eq_z_FN_layer_Reentrant_phases} -- \eqref{Full_Int_Eqb_FN_layer_Reentrant_phases} and shown in Fig.~\ref{fig_Phi0_H}b. Moreover, on phase diagram (Fig.~\ref{fig_Thresholds_EH}) there are regions of ambiguity, in which fixed electric field corresponds to different values of $\varkappa_H^{*}$. Decrease of orientational elasticity anisotropy $k$ leads to the increase of tricritical value $\varkappa_H^{*}$ of segregation parameter, besides, the width of ambiguity region $\eta$ does not depend on anisotropy constant $k$ (Fig.~\ref{fig_kappaH_TKT}a) which is consistent with expression \eqref{delta_FN_layer_Reentrant_phases1}. Decrease of parameter $b$ leads to the decrease of $\varkappa_H^{*}$ and the narrowing of ambiguity region (Fig.~\ref{fig_kappaH_TKT}b). The latter disappears only at $b=0$, that corresponds either to pure NLC, or to non-magnetic admixture. For parameters marked at the plane $(\varkappa, E)$ in Fig.~\ref{fig_kappaH_TKT}b by points $M_1=(0, 1.2)$,  $M_2=(4.0, 1.2)$, $P_1=(3.8, 2.0)$, $P_2=(3.8, 0.4)$, $P_3=(3.8, 0.25)$, the director orientation angle $\varphi_0$ in the middle of the layer is shown in Fig.~\ref{fig_Phi0_H}.

\subsection{Electric Freedericksz transitions in a magnetic field}

Let us now consider the Freedericksz transitions between uniform and nonuniform phases induced by electric field in the presence of a magnetic field. From expression \eqref{h_c_FN_layer_Reentrant_phases} and Fig.~\ref{fig_Thresholds_EH} it follows that the threshold electric field  $E_c$ of Freedericksz transition in FN is a single-valued function of $H$:
\begin{equation} \label{E_c_FN_layer_Reentrant_phases}
E_c=\sqrt{\pi^2+b Hs_E - H^2}, \quad s_E=2\sigma/(2\sigma + bH),
\end{equation}
so, the re-entrant orientational transitions can not be induced by the electric field. Moreover, magnetic field  $H$ should not exceed the critical field $H_c$  of the magnetic Freedericksz transition \cite{MakarovZakh_2010_PRE} in FN at  $E=0$, otherwise the Freedericksz transition in FN can not be induced by  electric field (Fig.~\ref{fig_Thresholds_EH}). If this requirement is satisfied, free energy \eqref{F_FN_layer_dimensionless_Reentrant_phases} close to  $E_c$  can be represented as a power series in $\varphi(z) = \varphi_0 \cos(\pi z)$ and $\psi(z) = s_E \varphi(z)$, where $\varphi_0 \ll 1$. The fourth-order Landau expansion of the free energy takes the form
\begin{equation} \label{F_criticalE_FN_layer_Reentrant_phases}
 \widetilde{F}  = \widetilde{F}_0  + \frac{\alpha_E}{2} \left(E_c - E\right)  \varphi_0^{2} +\frac{\beta_E}{4} \varphi_0^{4} + \dots,
\end{equation}
where
\begin{align*}
\widetilde{F}_0 &= -b H, \quad \alpha_E = E_c, \\
 \beta_E &= \frac{1}{16 \varkappa} \left[3 b H s_E^{2} \left(2-s_E\right)^2  +4\pi^2 k\right] \left(\varkappa-\varkappa_E^{*}\right).
\end{align*}
Here we introduce
\begin{equation} \label{kappaE_critical_FN_layer_Reentrant_phases}
 \varkappa_E^{*} = \frac{\left(b H s_E \right)^2}{3 b H  s_E^2  \left(2-s_E\right)^2  +4\pi^2 k}.
\end{equation}
Minimization of free energy \eqref{F_criticalE_FN_layer_Reentrant_phases} with respect to $\varphi_0$ gives the expression for the director rotation angle in the middle of the layer
\begin{equation} \label{phi0_close_TKT1_FN_layer_Reentrant_phases}
 \varphi_0 = \pm \sqrt{\gamma_E \frac{\,E-E_c}{\varkappa-\varkappa_E^{*}}}, \quad  \gamma_E= \displaystyle \frac{16 \varkappa E_c}{3b H s_E^{2} \left(2-s_E\right)^2  +4\pi^2 k} >0.
\end{equation}
Formula \eqref{phi0_close_TKT1_FN_layer_Reentrant_phases} shows that electric Freedericksz transition in a magnetic field also demonstrates tricritical behavior. For $\varkappa \geq \varkappa_E^{*}$ (weak segregation)  orientational transition is the second-order transition, and for  $\varkappa < \varkappa_E^{*}$ (strong segregation) --- the transition of the first order. The difference of Freedericksz transitions in FN induced by the electric field in the presence of magnetic field from magnetic Freedericksz transitions in electric field is that these transitions can not be re-entrant transitions, because critical field $E_c$  is a single-value function \eqref{E_c_FN_layer_Reentrant_phases} of a magnetic field $H$. At the zero magnetic field, $H=0$, electric orientational transitions in FN are the second-order, \cite{deGennesBook_en}  because $\varkappa_E^{*} = 0$.

\begin{figure}[!htb]
\begin{minipage}[!htb]{1.0\linewidth}
\center{\includegraphics[width=0.65\linewidth]{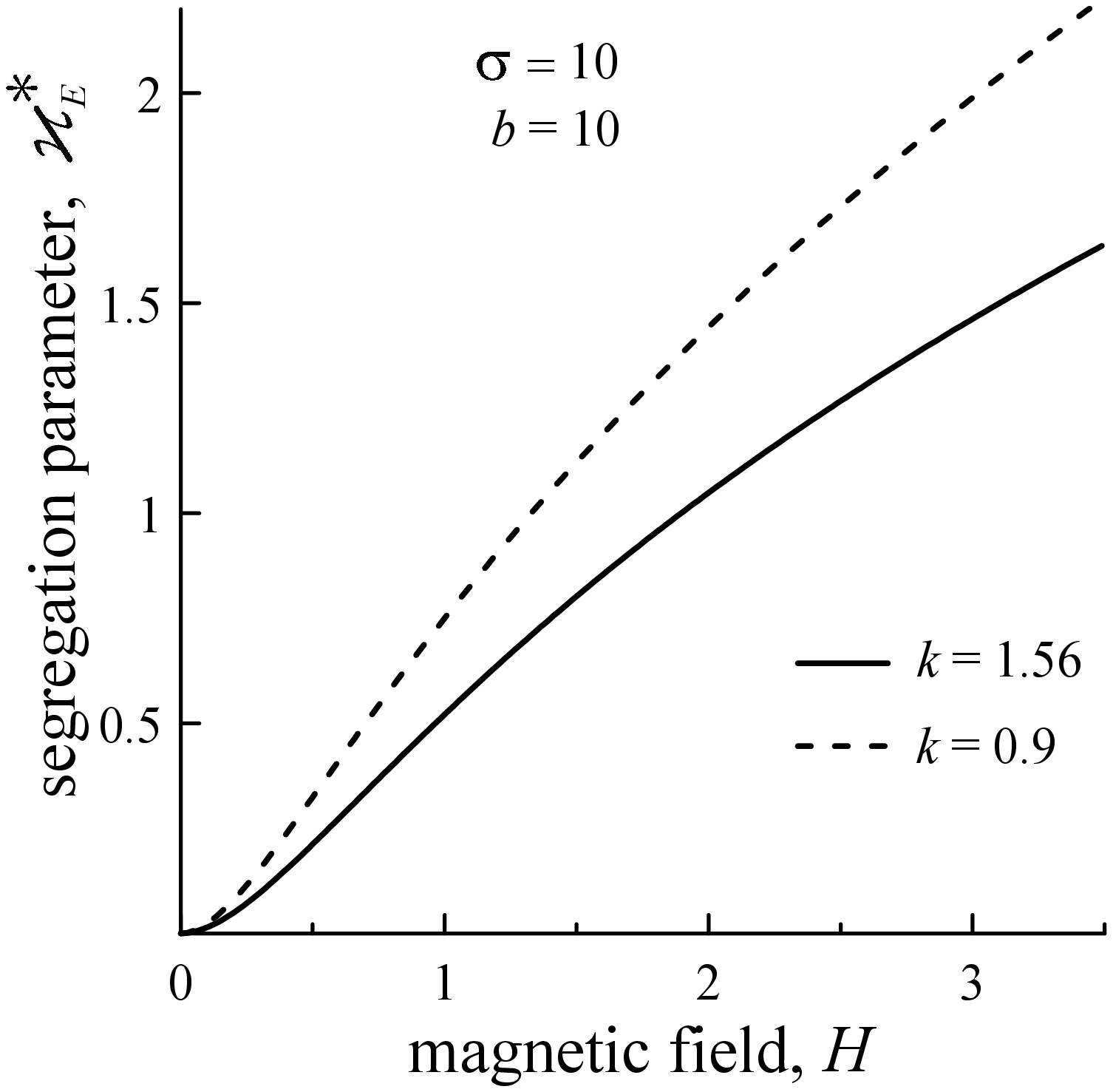} \\ (a)}
\end{minipage}
\vfill
\begin{minipage}[!htb]{1.0\linewidth}
\center{\includegraphics[width=0.65\linewidth]{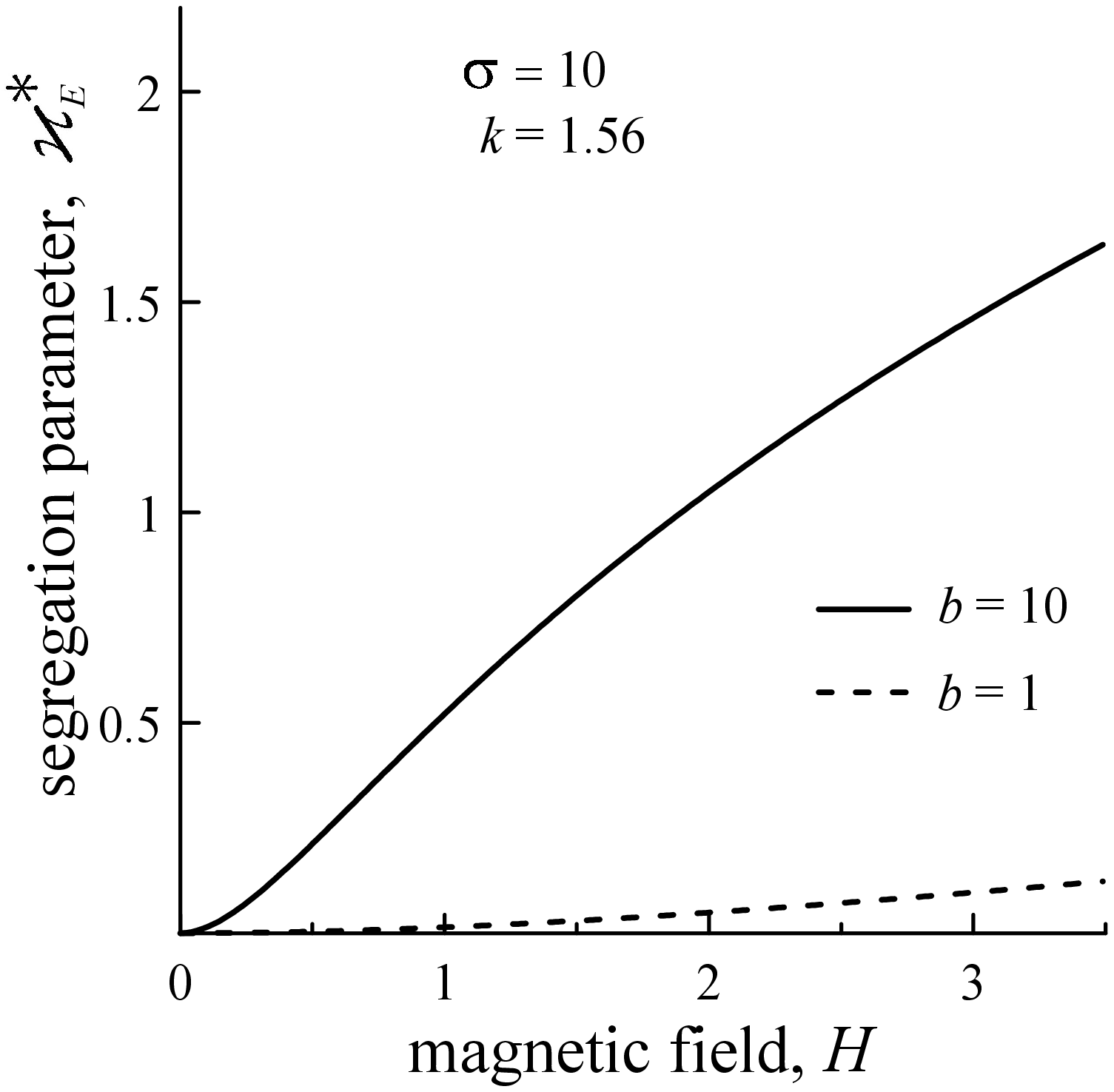} \\ (b)}
\end{minipage}
\caption{Tricritical value $\varkappa_E^{*}$ of the segregation parameter  as a function of magnetic field $H$ for different  values of (a) orientational elasticity anisotropy constant $k$ and (b) parameter $b$, which characterizes the modes of magnetic field influence on ferronematic}
\label{fig_kappaE_TKT}
\end{figure}

Tricritical value $\varkappa_E^{*}$  of the segregation parameter \eqref{kappaE_critical_FN_layer_Reentrant_phases} for the electric Freedericksz transition in FN \eqref{phi0_close_TKT1_FN_layer_Reentrant_phases} depends on the magnetic field $H$. As the
magnetic field increases the tricritical segregation parameter  increases  as shown in Fig.~\ref{fig_kappaE_TKT}.
Thus, if at zero magnetic field the electric Freedericksz transition in FN is a second-order transition (\textit{i.e.},  $\varkappa > \varkappa_E^{*}$), the magnetic field will lead to the increase of $\varkappa_E^{*}$ so that  $\varkappa < \varkappa_E^{*}$ and the electric Freedericksz transition in a magnetic field will become the first order (Fig.~\ref{fig_Phi0_E1}).

Moreover, decrease of orientational elasticity anisotropy $k$ increases the tricritical value of segregation parameter $\varkappa_E^{*}$ (Fig.~\ref{fig_kappaE_TKT}a), whereas decrease of parameter $b$, per contra, leads to the decrease of $\varkappa_E^{*}$ (Fig.~\ref{fig_kappaE_TKT}b).

\subsection{The field of the first-order equilibrium Freedericksz transition }

As stated above, magnetic orientational transition between uniform and nonuniform states of FN in the electric field for  $\varkappa < \varkappa_H^{*}$ or electric transition in a magnetic field for $\varkappa < \varkappa_E^{*}$  are the transitions of the first order. Therefore equilibrium transition fields  is not the same as $E_c$ or $H_c$. Let us determine magnetic $H_t$ and electric $E_t$ fields of equilibrium Freedericksz transitions of the first order. These fields are determined from the equality of the free energies of perturbed \eqref{F_FN_layer_dimensionless_Reentrant_phases} and unperturbed states of ferronematic
\begin{equation} \label{H_t_FN_layer_Reentrant_phases}
\widetilde{F} = \widetilde{F}_0,
\end{equation}
where the free energy of unperturbed state
\begin{equation} \label{F0_FN_dimensionless_pi/2_Reentrant_phases}
\widetilde{F}_0 \equiv \widetilde{F}|_{\varphi=\psi=0} = - bH.	
\end{equation}
Using eqn \eqref{Int2_FN_layer_Reentrant_phases} expression \eqref{H_t_FN_layer_Reentrant_phases} is reduced to
\begin{equation} \label{eqs_H_t_FN_layer_Reentrant_phases}
\begin{split}
   2 \int_{0}^{\varphi_0}\mathcal{R}^{-1} (\varphi, \psi) \, \mathcal{K}^2(\varphi) d\varphi - \frac{1}{2} (H^{2} + E^2) \sin^{2} \varphi_0 \\
   + \varkappa \left[ \ln eQ - g_0  \right] +  b H = 0,
 \end{split}
\end{equation}
where  $e$ is the base of the natural logarithm, whereas functions $\mathcal{K}(\varphi)$, $\mathcal{R}(\varphi, \psi)$, $g(\varphi, \psi)$, and normalization integral $Q$ are determined by expressions \eqref{K_function_FN_layer_Reentrant_phases}, \eqref{R_function_FN_layer_Reentrant_phases}, \eqref{Full_Int_Eqb_FN_layer_Reentrant_phases}, and \eqref{Q_function_FN_layer_Reentrant_phases} respectively. This equation is solved together with the equations of orientational equilibrium \eqref{Eq_System_B_FN_layer_Reentrant_phases}, \eqref{Full_Int_Eqa_FN_layer_Reentrant_phases}, and \eqref{Full_Int_Eqb_FN_layer_Reentrant_phases}. For magnetic Freedericksz transition in a given electric field $E$ eqn \eqref{eqs_H_t_FN_layer_Reentrant_phases} allows to obtain $H_t$ (Fig.~\ref{fig_Phi0_H}), and in the case of electric Freedericksz transition at fixed magnetic field $H$ it determines $E_t$ (Fig.~\ref{fig_Phi0_E1}).

\section{\label{sec_Numerical_results_Reentrant_phases} Orientational, magnetic and optical properties of ferronematic}

\subsection{Magnetic Freedericksz transitions in an electric field}

\begin{figure}[!htb]
\begin{minipage}[!htb]{1.0\linewidth}
\center{\includegraphics[width=0.65\linewidth]{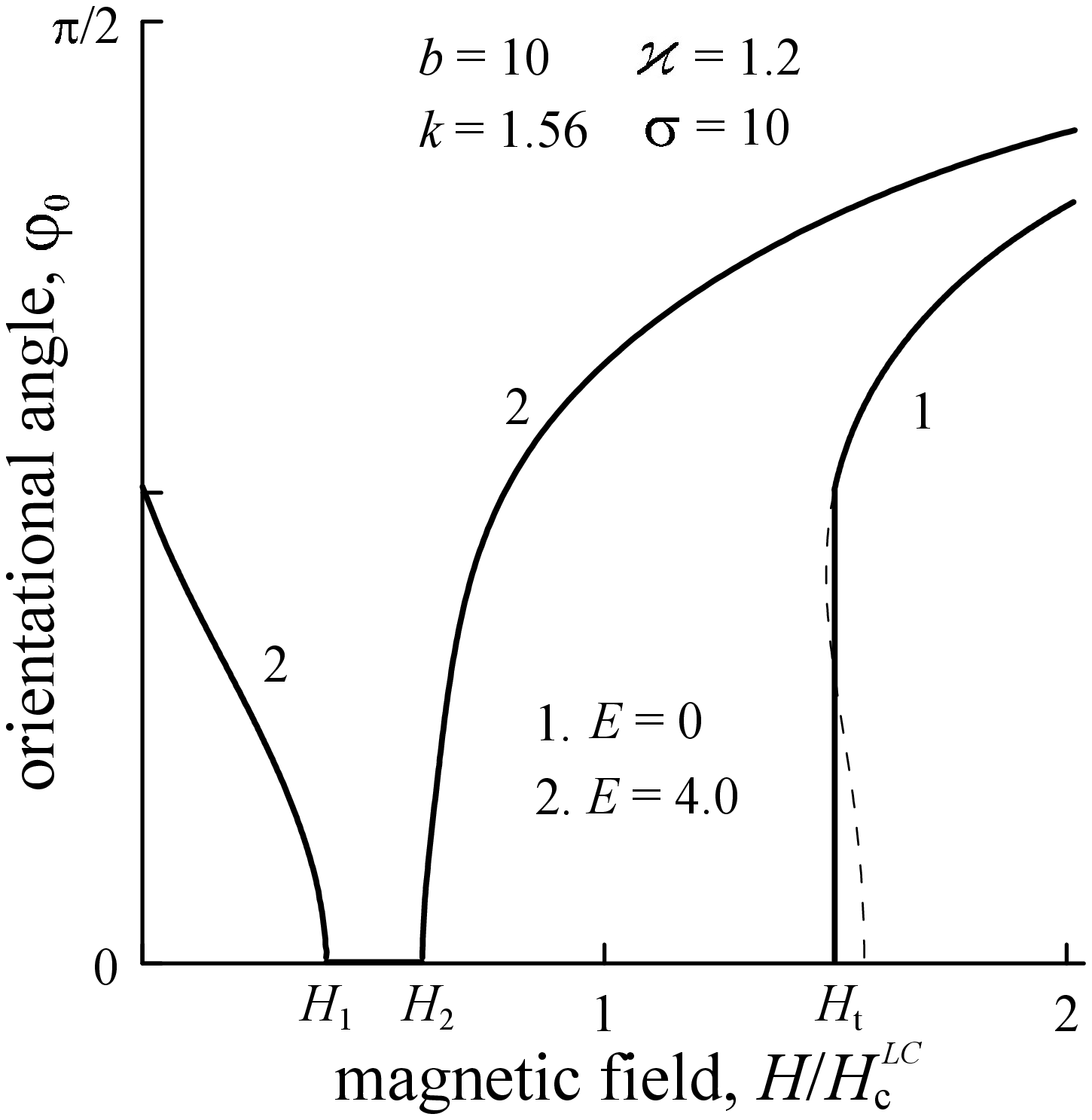} \\ (a)}
\end{minipage}
\hfill
\begin{minipage}[!htb]{1.0\linewidth}
\center{\includegraphics[width=0.65\linewidth]{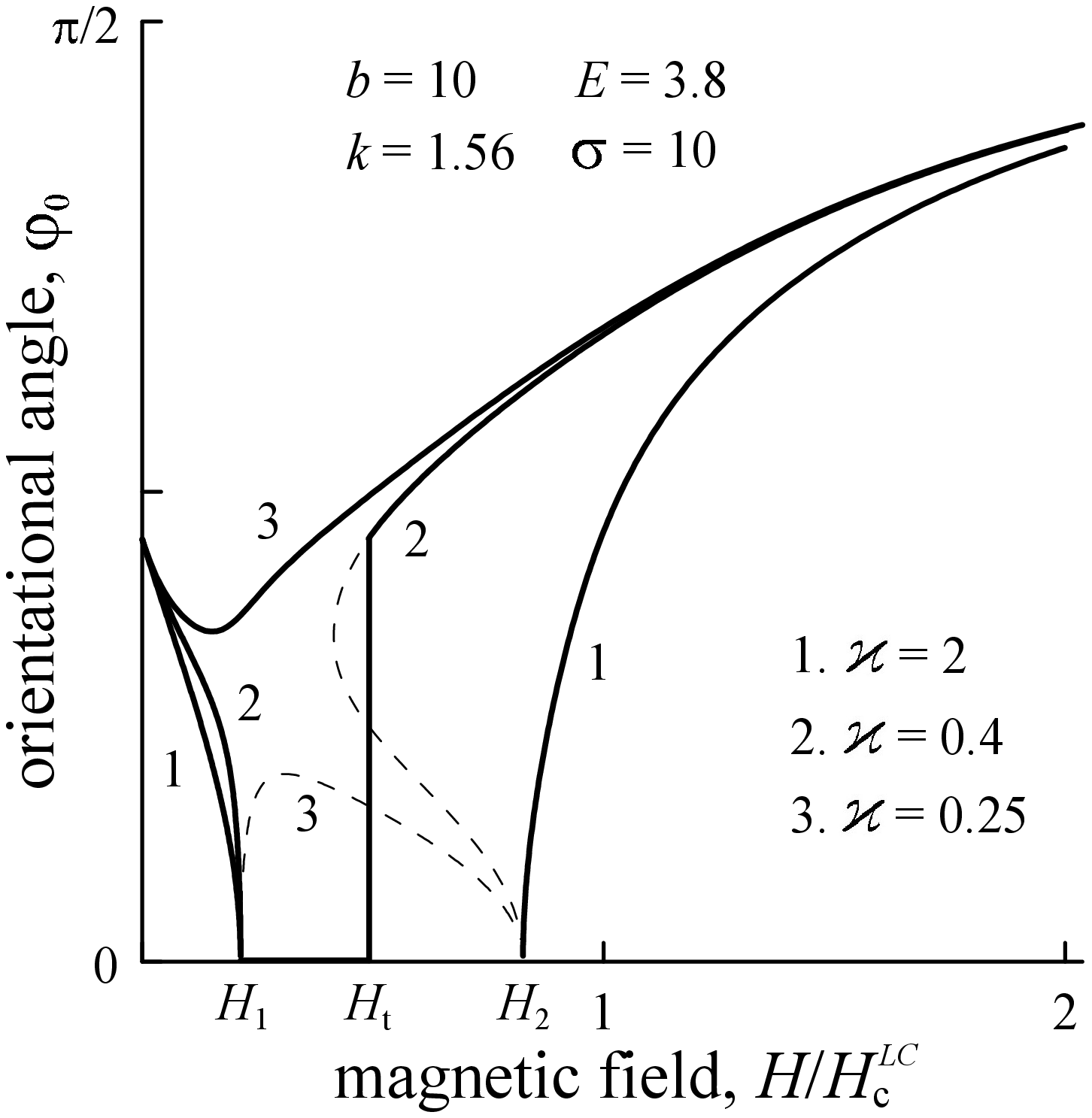} \\ (b)}
\end{minipage}
\caption{Director orientation angle $\varphi_0$ in the middle of FN layer  as a function of magnetic field $H$ for different values of (a) electric field $E$ ($H_1 = 1.25$, $H_2 = 1.90$, $H_t = 4.72$) and (b) segregation parameter $\varkappa$  ($H_1 = 0.67$, $H_2 = 2.59$, $H_t = 1.53$); dashed lines represent metastable and unstable states}
\label{fig_Phi0_H}
\end{figure}

At zero electric field, $\boldsymbol{\mathcal{E}}=0$, and a weak magnetic field $\boldsymbol{\mathcal{H}}$ which is applied perpendicular to the layer, the orientational structure of FN  is uniform and planar with orthogonal orientation of the director and magnetization ($\mathbf{n} \bot \mathbf{m}$). With the increase of the magnetic field this state becomes unstable and at a certain critical value the orientational structure of FN becomes deformed, \textit{i.e.}, magnetic Freedericksz transition takes place.  As shown in Ref. \cite{MakarovZakh_2010_PRE}, for weak segregation ($\varkappa > \varkappa_H^{*}$) magnetic Freedericksz transition is the second-order transition, for which $H_c$  \eqref{h_c_FN_layer_Reentrant_phases} plays role of critical field. At strong segregation ($\varkappa < \varkappa_H^{*}$) the orientational transition occurs as the first order phase transition (Fig.~\ref{fig_Phi0_H}a --- curve 1, Fig.~\ref{fig_kappaH_TKT}b --- point $M_1$). In this case the field $H_t$ of equilibrium transition is determined by eqn \eqref{eqs_H_t_FN_layer_Reentrant_phases}.

Application of electric field $\boldsymbol{\mathcal{E}} \parallel \boldsymbol{\mathcal{H}}$ may change the character of orientational transition. The increase of its strength not only decreases the transition field $H_c$ \eqref{h_c_FN_layer_Reentrant_phases}, but also decreases the tricritical value $\varkappa_H^{*}$ of segregation parameter  \eqref{kappaH_critical_FN_layer_Reentrant_phases}, that allows to satisfy the condition $\varkappa > \varkappa_H^{*}$. In this case orientational transition becomes the transition of the second order (Fig.~\ref{fig_Phi0_H}a --- right curve 2, Fig.~\ref{fig_kappaH_TKT}b --- point $M_2$). For electric field belongs to the interval $E^{LC}_c  \leq E \leq E^{LC}_c + \eta$ corresponding to ambiguity region of the phase diagram (Fig.~\ref{fig_kappaH_TKT}b), the increase of magnetic field leads to a sequence of re-entrant orientational phases: nonuniform phase ($H<H_1$) --- uniform phase ($H_1 \leq H \leq H_2$) --- nonuniform phase ($H>H_2$). For $E \geq E^{LC}_c$ at $H=0$ uniform texture of FN becomes unstable and electric Freedericksz transition takes place. Application of a magnetic field decreases the deviation of a system from uniform orientational state (decrease of director rotation angle inside the layer). Subsequent increase of $H$ returns the orientational structure of FN at $H=H_1$ into initial uniform state. This effect is due to a competition between quadrupole [$\sim \varepsilon_0 \varepsilon_a ( \mathbf{n} \cdot \boldsymbol{\mathcal{E}})^2$] mechanism of the electric field influence on NLC-matrix and the dipole [$ \sim \mu_0 M_s f \, \mathbf{m} \cdot \boldsymbol{\mathcal{H}}$] influence of magnetic field on magnetic moments of ferroparticles. When magnetic field is applied, the particles tend to align along the direction of the field, but since they are coupled [$ \sim \frac{w}{d} f  \, (\mathbf{n} \cdot \mathbf{m})^2$] to NLC-matrix, they return the NLC-matrix to the initial (uniform) state. This state, however, becomes unstable again at magnetic field  $H=H_2$, and  Freedericksz transition occurs once again. In this case magnetic quadrupole [$\sim  \mu_0 \chi_a ( \mathbf{n} \cdot \boldsymbol{\mathcal{H}})^2$] mechanism influencing the liquid-crystal matrix prevails over the magnetic dipole [$ \sim \mu_0 M_s f \, \mathbf{m} \cdot \boldsymbol{\mathcal{H}}$] mechanism. Further increase of magnetic field gives rise to the increase of FN director rotation angle.

Figure~\ref{fig_Phi0_H}b represents the director rotation angle $\varphi_0$ in a center of FN layer as a function of magnetic field $H$ for different values of the segregation parameter $\varkappa$ at fixed $E$. Solid lines represent stable solutions corresponding to minimal free energy of FN, dashed lines --- metastable and unstable solutions. For  $\varkappa > \varkappa_H^{*}$ (Fig.~\ref{fig_Phi0_H}b --- curve 1, Fig.~\ref{fig_kappaH_TKT}b --- point $P_1$) as the magnetic field $H$ increases the transition from nonuniform state to uniform at $H=H_1$ and  subsequent transition to nonuniform state at $H=H_2$ occur continuously, \textit{i.e.}, they are the second-order phase transitions. With a decrease of segregation parameter  ($\varkappa < \varkappa_H^{*}$, Fig.~\ref{fig_Phi0_H}b --- curve 2, Fig.~\ref{fig_kappaH_TKT}b --- point $P_2$) the transition from nonuniform state to uniform at $H=H_1$ remains the second order, while subsequent orientational transition to nonuniform state has a jump at $H=H_t$ \eqref{eqs_H_t_FN_layer_Reentrant_phases}, indicative of the first-order transition. Under further decrease of segregation parameter ($\varkappa < \varkappa_H^{*}$, Fig.~\ref{fig_Phi0_H}b --- curve 3, Fig.~\ref{fig_kappaH_TKT}b --- point $P_3$) at a given value of electric field the re-entrant transitions are absent.

\subsection{Electric Freedericksz transition in a magnetic field}

\begin{figure}[!htb]
\center{\includegraphics[width=0.65\linewidth]{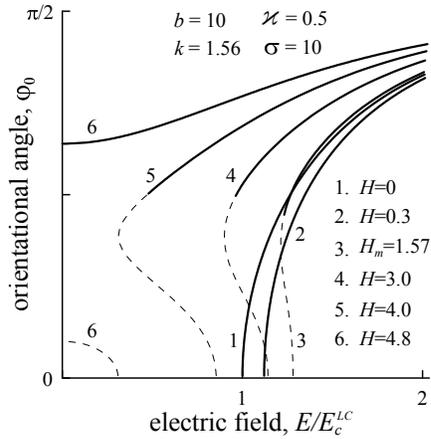}}
\caption{Director orientation angle $\varphi_0$ in the middle of FN layer as a function of electric field $E$; Stable branches of solutions are represented by solid lines, whereas metastable and unstable --- by dashed lines}
\label{fig_Phi0_E1}
\end{figure}

At zero magnetic field, $H=0$, the embedded magnetic particles behave themselves as a passive admixture. In this case the electric field $E$ applied to FN affects only the NLC-matrix, that is why electric orientational transition in FN (Fig.~\ref{fig_Phi0_E1}, curve 1) coincides with classical second-order Freedericksz transition in NLC.  \cite{deGennesBook_en}

Presence of a magnetic field  $\boldsymbol{\mathcal{H}} \parallel \boldsymbol{\mathcal{E}}$  can lead to a change of the critical field of electric Freedericksz transition and to the change of transition character from the second order to the first. In weak magnetic fields for $\varkappa > \varkappa_E^{*}$ (Fig.~\ref{fig_kappaE_TKT}) the transition field increases but transition character remains unchanged. The growth of a magnetic field up to $H=H_m$ (where $H_m$ is the magnetic field corresponding to the maximum of electric field $E_m=E_c^{LC}+\eta$ on phase diagram Fig.~\ref{fig_Thresholds_EH}) increases the transition field for $\varkappa < \varkappa_E^{*}$ (see Fig.~\ref{fig_kappaE_TKT}) this transition becomes the transition of the first order (Fig.~\ref{fig_Phi0_E1}, curve 3). With further increase of the magnetic field the character of electric Freedericksz transition in FN does not change but the value of the critical field  decreases (Fig.~\ref{fig_Phi0_E1}, curves 4--6). Such nonmonotonic behavior of electric Freedericksz transition in FN under magnetic field accompanied by the change of orientational transition character is due to the fact that external magnetic field $\boldsymbol{\mathcal{H}} \parallel \boldsymbol{\mathcal{E}}$ affects not only the director $\mathbf{n}$  [quadrupole magnetic mechanism of influence $\sim \mu_0 \chi_a (\mathbf{n} \cdot \boldsymbol{\mathcal{H}})^2$], but also the magnetization $\mathbf{m}$  [dipole mechanism $\sim \mu_0 M_s f \, \mathbf{m} \cdot \boldsymbol{\mathcal{H}}$].  Coupling between the director and magnetization [term $F_5$ in eqn \eqref{F_FN_layer_Reentrant_phases}] and their mutual orthogonal orientation lead to the competition of these mechanisms. In weak magnetic fields the dipole mechanism, responsible for orientation of magnetization, prevails over the quadrupole one, thus increasing the transition field. In strong magnetic fields prevails quadrupole mechanism, responsible for the director behavior, which decreases the critical field. The change of transition character is connected with the dependence of tricritical value of segregation parameter $\varkappa_E^{*}$ \eqref{kappaE_critical_FN_layer_Reentrant_phases} on magnetic field, that allows to make $\varkappa_E^{*}$ either more, or less $\varkappa_E$ (Fig.~\ref{fig_kappaE_TKT}).

\subsection{Ferronematic magnetization}

Magnetization of ferronematic $\boldsymbol{\mathcal{M}} = M_s f \mathbf{m}$ is determined by the combination of concentrational $f$ and orientational $\mathbf{m}$ distributions of magnetic particles in a layer. Let us define the reduced magnetization of ferronematic \cite{ZakhShavkunov_2000_JMMM}
\begin{equation*}
\mathbf{M} = \boldsymbol{\mathcal{M}}/(M_s \overline{f})=g \mathbf{m} = \textbf{[} - g(z) \sin\psi(z), \, 0, \, g(z)  \cos \psi(z)\textbf{]},
\end{equation*}
which taking into account the concentrational distribution $g(z)$ takes the form
\begin{equation}\label{Mag_FN_layer_Reentrant_phases}
\mathbf{M} = \mathbf{m}  Q \exp{\left\{\displaystyle \frac{b H}{\varkappa} \cos \psi  - \frac{\sigma}{\varkappa} \sin^2 (\varphi - \psi)\right\}}.
\end{equation}

\begin{figure}[!htb]
\begin{minipage}[!htb]{1.0\linewidth}
\center{\includegraphics[width=0.65\linewidth]{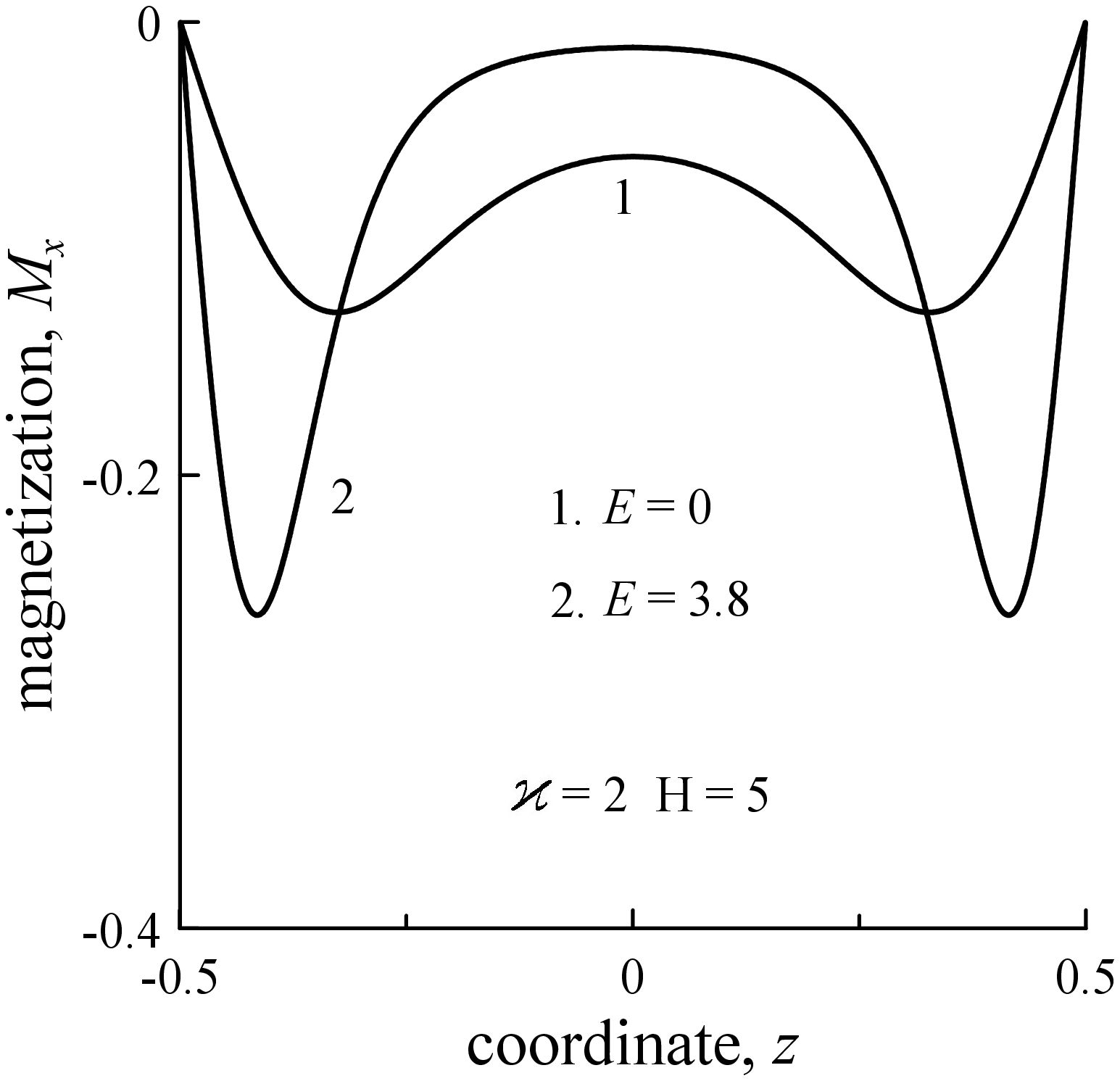} \\ (a)}
\end{minipage}
\vfill
\begin{minipage}[!htb]{1.0\linewidth}
\center{\includegraphics[width=0.65\linewidth]{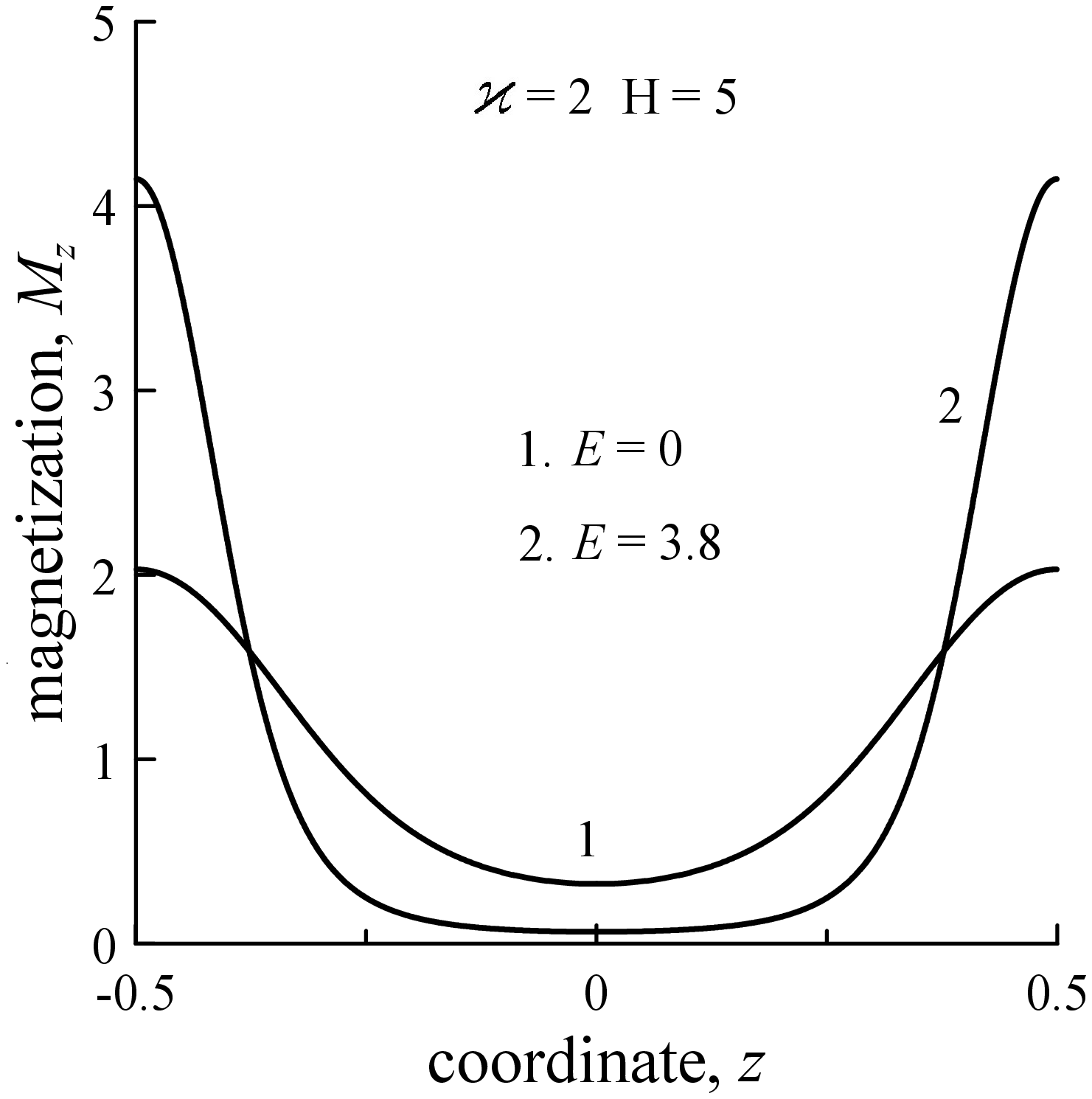} \\ (b)}
\end{minipage}
\caption{Reduced magnetization $\mathbf{M}$ of ferronematic layer for $\varkappa=2$, $H=5$, $\sigma =10$, $k = 1.56$, $b = 10$}
\label{fig_mag}
\end{figure}

Figure~\ref{fig_mag} represents the components of reduced magnetization $\mathbf{M}$ in FN layer for two values of electric field $E$. Chosen value of magnetic field $H=5$ is greater than the field $H_c = 4.91$ of Freedericksz transition in ferronematic at zero electric field. \cite{MakarovZakh_2010_PRE} In this case, even in the absence of electric field, there are orientational distortions and nonuniformity of magnetic particles distribution  in the FN layer (segregation effect). At $E=0$ and $H\neq0$ (Fig.~\ref{fig_mag}b, curve 1) due to increasing of concentration of magnetic particles near the boundaries of the layer,  $M_z$ is maximal at the boundaries of the layer, while the depleted of magnetic particles central part of the FN layer is practically not magnetized. Due to rigid planar conditions of the director coupling to the boundaries of the layer,  $|M_x|$ attains its maximum not on the boundaries, but in-between the center of the layer and its boundaries (Fig.~\ref{fig_mag}a, curve 1).

Application of electric field  (Fig.~\ref{fig_mag}, curves 2) contributes to further distortion of FN orientational structure and decreasing of magnetic particles concentration in the center of the layer, which leads to further increase of  $M_z$  on the layer boundaries and to the growth of  $|M_x|$ near them.

\begin{figure}[!htb]
\begin{minipage}[!htb]{1.0\linewidth}
\center{\includegraphics[width=0.65\linewidth]{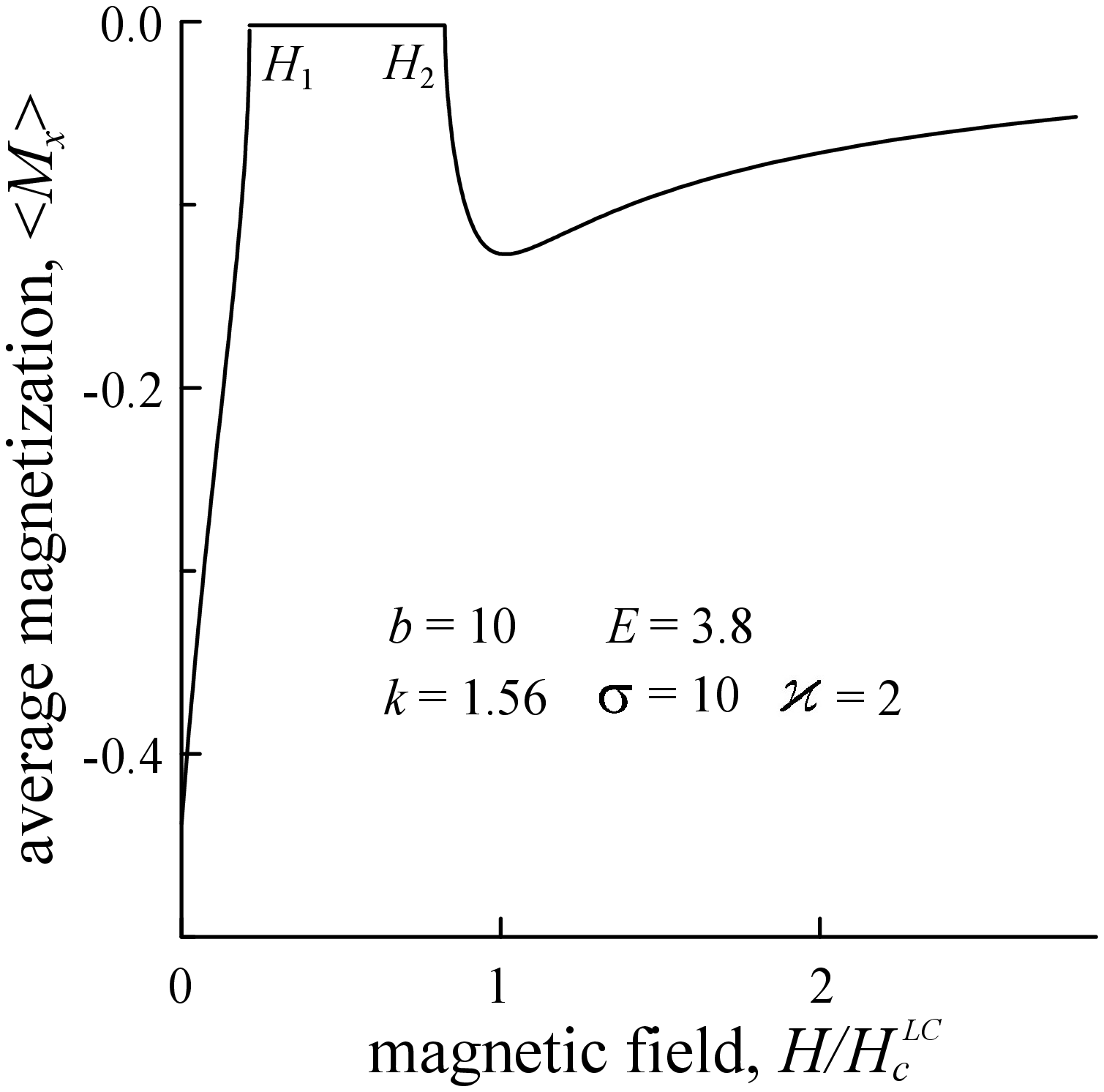} \\ (a)}
\end{minipage}
\vfill
\begin{minipage}[!htb]{1.0\linewidth}
\center{\includegraphics[width=0.65\linewidth]{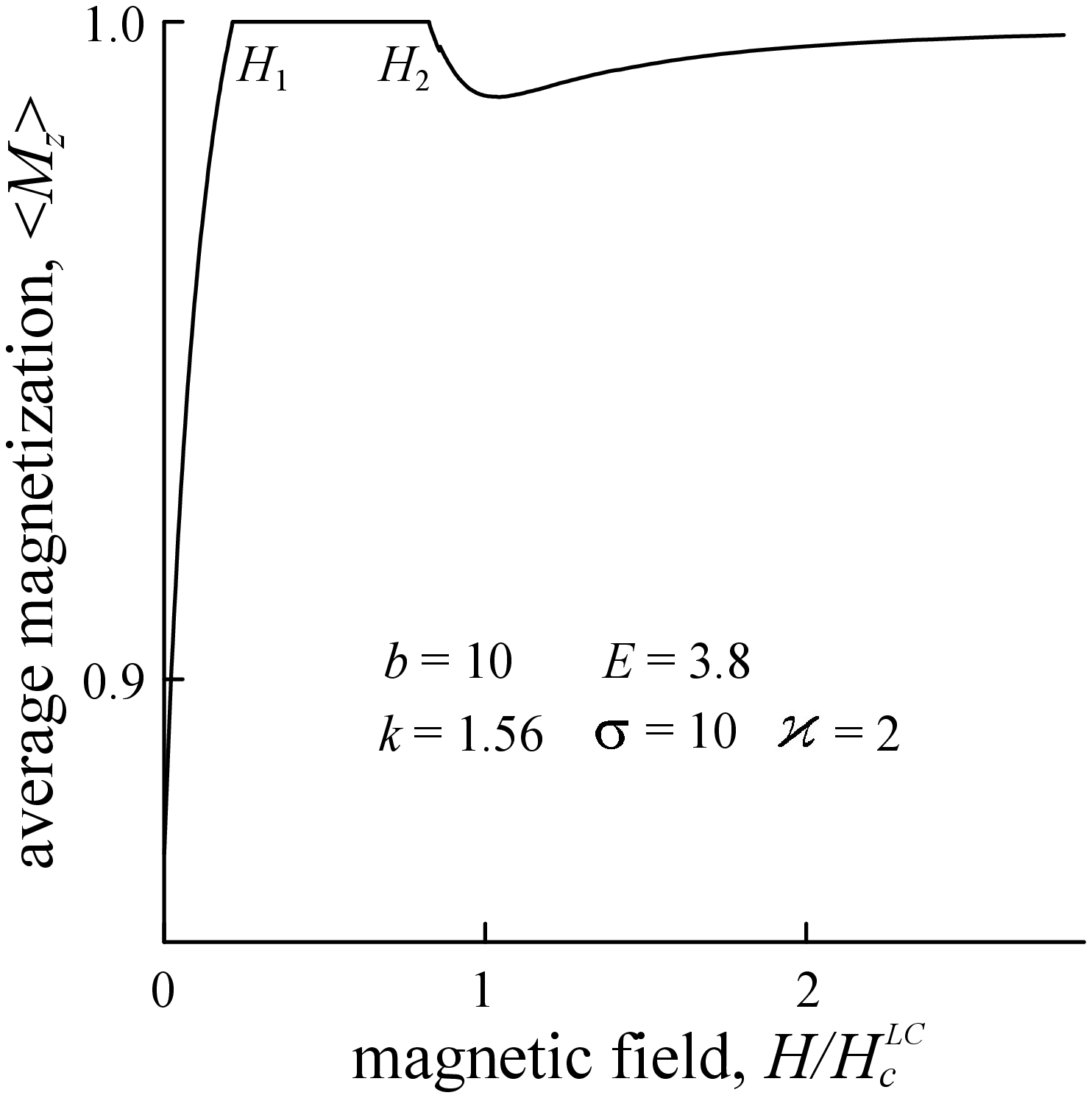} \\ (b)}
\end{minipage}
\caption{Average value of a reduced magnetization $\langle \mathbf{M} \rangle$ of a ferronematic as a function of magnetic field $H$ for $\varkappa=2$, $E=3.8$, $\sigma =10$, $k = 1.56$, $b = 10$}
\label{fig_mag_avg}
\end{figure}

Formula \eqref{Mag_FN_layer_Reentrant_phases}  enables us to obtain the average magnetization across  the layer:
\begin{align}\label{avg_Mag_FN_layer_Reentrant_phases}
&\langle M_x \rangle = -2 \int_{0}^{\varphi_0} g \sin \psi \; \sqrt{\mathcal{R}(\varphi, \psi) } \, d \varphi, \nonumber \\
& \langle M_y \rangle = 0, \quad \langle M_z \rangle = 2 \int_{0}^{\varphi_0} g \cos  \psi \; \sqrt{\mathcal{R}(\varphi, \psi)} \, d \varphi,
\end{align}
which is shown in Fig.~\ref{fig_mag_avg} as a function of $H$. In the absence of magnetic and electric fields the FN orientational texture and distribution of magnetic particles in a layer are uniform: $\mathbf{n} = (1, 0, 0)$, $\mathbf{m} = (0, 0, 1)$ and $g(z)=1$. In this case average magnetization $\langle\mathbf{M}\rangle$ has only $z$-component $\langle M_z \rangle = 1$. Application of an electric field $E>E_c^{LC}$ leads to the distortion of FN orientational structure (electric Freedericksz transition) and, as a consequence, to the change of average values of the suspension magnetization components. For numerical simulations, electric field  has been chosen from the re-entrant phase transition interval $E^{LC}_c  \leq E \leq E^{LC}_c + \eta$ (see Fig.~\ref{fig_Thresholds_EH}). As is seen from Fig.~\ref{fig_mag_avg}, application of a magnetic field and the increase of its strength returns FN into initial orientational state at $H=H_1$ with  $\langle M_x \rangle =0$  and $\langle M_z \rangle =1$.  Further increase of the field $H$  leads to the second Freedericksz transition from uniform to nonuniform phase at $H=H_2$. At first, there is a decrease of  $\langle M_z \rangle$ and  $\langle M_x \rangle$ because of the coupling between the director and magnetization, and then, there appears the asymptotic convergence of average magnetization values to their values in the undisturbed ferronematic: $\langle M_x \rangle = 0$ and $\langle M_z \rangle=1$.

Near the transition from uniform  to nonuniform phase the reduced magnetization \eqref{Mag_FN_layer_Reentrant_phases} can be represented  as a powers  series in  $\varphi(z)$  and $\psi(z)$ \eqref{PhiPsi_FN_layer_Reentrant_phases},
where $\varphi_0 \ll 1$  is determined by relation  \eqref{phi0_close_TKT_FN_layer_Reentrant_phases}. In the case of re-entrant phase transitions at $E \neq 0$ (see Fig.~\ref{fig_Thresholds_EH}), the Freedericksz transition (direct and re-entrant) takes place at  $H$ equal to $H_1$ and $H_2$. Then the first nonvanishing terms of expansion near $H_c$ for averaged over the layer the components of reduced magnetization  \eqref{avg_Mag_FN_layer_Reentrant_phases} have the form
\begin{equation}
\langle M_x \rangle = - \frac{2}{\pi} s_H \varphi_0, \qquad \langle M_z \rangle = 1 - \frac{1}{4} s_H^2 \varphi_0^2,
\end{equation}
from which the nonzero components of the FN dipole part of magnetic susceptibility tensor  $\chi_{ij}^d= \partial \langle M_i\rangle / \partial H_j$ can be written as follows

\begin{equation}
\chi_{xz}^d = - \frac{s_H}{\pi} \sqrt{\frac{\gamma_H}{H-H_c}}, \qquad \chi_{zz}^d = - \frac{1}{4} s_H^2 \gamma_H,
\end{equation}
where $\gamma_H$ is determined by relation  \eqref{phi0_close_TKT_FN_layer_Reentrant_phases}.

\subsection{Optical phase lag}

Distortion of ferronematic orientational structure can be observed experimentally by measuring optical phase difference of ordinary and extraordinary rays after they pass through the ferronematic layer. Let us find out how phase difference changes with increasing magnetic field using the following formula:   \cite{BlinovBook_1994_en}
\begin{equation} \label{Optical_lag_z_FN_layer_Reentrant_phases}
   \delta = \frac{2 \pi L}{\lambda} \int_{-1/2}^{1/2}  \left[n_{eff}(z) - n_{o}\right] dz,
\end{equation}
where $\lambda$ is the wavelength of the transmitted beam of monochromatic light incident normally to the layer surface, $n_{eff}(z)$ is the effective refractive index, defined by the relation
\begin{equation*}
   \frac{1}{n_{eff}^2(z)} =   \frac{\sin^2\varphi(z)}{n_{o}^2} + \frac{\cos^2\varphi(z)}{n_{e}^2},
\end{equation*}
where $n_{o}$ and $n_{e}$ are the refractive indices of the ordinary and the extraordinary rays, respectively.

\begin{figure}[!htb]
\begin{center}
\includegraphics[width=0.65\linewidth]{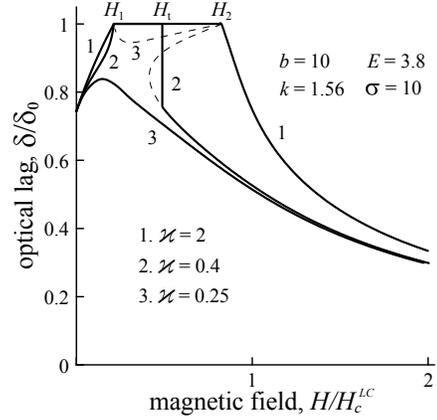}
\caption{Reduced phase lag $\delta/\delta_0$ in ferronematic layer as a function of magnetic field $H$ for  $E =3.8$, $\sigma =10$, $k = 1.56$, $b = 10$, $\xi = 0.2$. Here  $\varkappa_H^{*}=1.30$, $H_1 = 0.67$, $H_2 = 2.59$, $H_t = 1.54$; solid lines correspond to stable states, whereas dashed lines --- metastable and unstable}
\label{fig_lag}
\end{center}
\end{figure}

Using condition \eqref{Int2_FN_layer_Reentrant_phases}, reduced phase lag \eqref{Optical_lag_z_FN_layer_Reentrant_phases} in ferronematic distorted by applied magnetic and electric fields can be written as
\begin{equation} \label{Optical_lag_phi_FN_layer_Reentrant_phases}
   \frac{\delta}{\delta_0} = 2 \int_{0}^{\varphi_0}   \frac{\left( 1-\xi +\sqrt{1-\xi}\right)\cos^{2}\varphi}{ 1-\xi\cos^{2}\varphi +\sqrt{1-\xi\cos^{2}\varphi}}   \sqrt{ \mathcal{R}(\varphi, \psi)} \, d\varphi,
\end{equation}
here we introduce the parameter $\xi = \left(n_{e}^2 - n_{o}^2\right)/n_{e}^2$ and the phase lag in the absence of magnetic and electric fields $\delta_0 = 2 \pi L\left (n_{e} - n_{o}\right)/\lambda$. In formula \eqref{Optical_lag_phi_FN_layer_Reentrant_phases} the angles $\varphi_0$, $\psi_0$ and distribution function $g$ are given by eqns \eqref{Eq_System_B_FN_layer_Reentrant_phases}, \eqref{Full_Int_Eqa_FN_layer_Reentrant_phases}, and \eqref{Full_Int_Eqb_FN_layer_Reentrant_phases}, whereas the function $\mathcal{R}(\varphi, \psi)$ --- by relation \eqref{R_function_FN_layer_Reentrant_phases}. The reduced phase lag $\delta/\delta_0$ \eqref{Optical_lag_phi_FN_layer_Reentrant_phases} in ferronematic layer as a function of magnetic field $H$ is shown in Fig.~\ref{fig_lag}.

Note, that phase lag at $E=0$ has been studied previously in Ref. \cite{MakarovZakh_2010_PRE}. In the presence of an electric field for weak magnetic segregation ($\varkappa > \varkappa_H^{*}$, Fig.~\ref{fig_lag}, curve 1) a sequence of re-entrant magnetic transitions takes place: nonuniform phase ($H < H_1$)  --- uniform phase ($H_1 \leq H \leq H_2$)  --- nonuniform phase ($H > H_2$). The phase lag varies continuously with the increase of magnetic field --- both transitions are of the second order. In the interval $H_1 \leq H \leq H_2$  ferronematic layer has maximal birefringence. Phase lag tends asymptotically to zero at $H \gg H_c^{LC}$ because of rigid coupling of the director with the layer boundaries. With the decrease of segregation parameter ($\varkappa < \varkappa_H^{*}$, Fig.~\ref{fig_lag}, curve 2) phase lag decreases discontinuously at $H=H_t$ (the first-order transition), and then decreases with increasing field. The transition from nonuniform to uniform state at $H=H_1$ is still the transition of the second order. For smaller values of segregation parameter ($\varkappa \ll \varkappa_H^{*}$, Fig.~\ref{fig_lag}, curve 3) there is no re-entrant orientational transitions. Application of a magnetic field initially increases the phase lag, which never reaches maximum value $\delta =\delta_0$, because FN orientational structure does not return to initial uniform state. Further increase of magnetic field leads to the decrease of phase lag.

\section{\label{sec_Conclusion}  Conclusion}

In the present paper we have studied the influence of magnetic and electric fields on a suspension of magnetic particles in nematic liquid crystal (ferronematic). Both fields are able to induce the orientational Freedericksz transitions in ferronematics independently of one another. However, the response of a suspension on these fields is different. An external magnetic field affects both the orientation of magnetic particles (dipole mechanism) and the liquid crystal structure (quadrupole mechanism). Coupling the director with magnetization leads to competition between these mechanisms of influence. In contrast to magnetic field an electric field can directly affect one of the subsystems --- liquid crystal. We have shown that in certain electric field range the magnetic field can induce a sequence of re-entrant orientational transitions in ferronematic: nonuniform phase --- uniform phase --- nonuniform phase. This phenomenon is caused by the interplay between the dipole (ferromagnetic) and quadrupole (dielectric and diamagnetic) mechanisms of the field influence on a ferronematic structure.  We have found that these re-entrant Freedericksz transitions exhibit tricritical behavior, \textit{i.e.}, they can be of the first or the second order. The character of the transitions depends on a segregation degree and external fields. Moreover, electric field can change the character of the magnetic Freedericksz transition in ferronematic from the first order to the second, but not \textit{vice versa}.

We have shown that re-entrant Freedericksz transitions in ferronematic can not be induced by electric field at any fixed values of a magnetic field, since the threshold electric field for the orientational transition is the single-valued function of a magnetic field. The presence of a magnetic field can change the type of electric Freedericksz transition in ferronematic from the second order to the first only. In addition, we reveal a nonmonotonic behavior of the threshold electric field for this equilibrium transition.

 \section{Acknowledgement}
This work was partially supported by Russian Foundation for Basic Research, grant No. 10-02-96030.

\footnotesize{
\bibliography{Bib_Makarov} 
\bibliographystyle{rsc} 
}

\end{document}